\documentclass[fleqn,10pt]{wlscirep}
\usepackage[utf8]{inputenc}
\usepackage[T1]{fontenc}
\usepackage{bm}
\usepackage{amsmath}

\title{Can LSTM outperform volatility-econometric models?}

\author[1,*]{German Rodikov}
\author[2]{Nino Antulov-Fantulin}
\affil[1]{Affiliation, SNS, Pisa}
\affil[2]{Affiliation, ETH, Zürich}
\affil[*]{e-mail: german.rodikov@sns.com, germanrodikov@gmail.com}

\begin{abstract}

Volatility prediction for financial assets is one of the essential questions for understanding financial risks and quadratic price variation. However, although many novel deep learning models were recently proposed, they still have a "hard time" surpassing strong econometric volatility models. Why is this the case? The volatility prediction task is of non-trivial complexity due to noise, market microstructure, heteroscedasticity, exogenous and asymmetric effect of news, and the presence of different time scales, among others. In this paper, we analyze the class of long short-term memory (LSTM) recurrent neural networks for the task of volatility prediction and compare it with strong volatility-econometric models.

\end{abstract}

\begin{document}
\flushbottom
\maketitle
\thispagestyle{empty}

\section{Introduction}

Although the first stochastic mathematical models of price changes, also known as Brownian motion or Wiener process, were proposed a long time ago ~\cite{bachelier2011louis, jarrow2004short}, they still did not fully describe empirical facts ~\cite{cont2001empirical} connected to the quadratic variation and thus volatility of asset prices.
In financial markets, volatility is associated with the level of price fluctuations.

Volatility changes over time; high volatility means high risk and sharp price fluctuations, while low volatility refers to smooth price changes \cite{Engle_1982,  Bollerslev_1986}. For derivative financial instruments, the price is directly referred to as underlying assets (implied) volatility. Thus, it plays a crucial role in these financial instruments. Options and Derivatives have increased in volume over the past decades \cite{poon2003forecasting}. Derivative contracts are written based on volatility measurement. In other words, volatility is the underlying asset. Thus, we can understand why so many researchers and practitioners focus on the volatility model and mostly on forecasting volatility models.

Some models are characterizing the volatility from a conditional process perspective, for example, ARCH family \cite{Engle_1982,  Bollerslev_1986}. The main idea of the conditional process perspective is the use of conditional variance, the values of which change over time, while the unconditional variance can remain relatively constant. The drawback of autoregression models for conditional heteroscedasticity is accuracy performance. Moreover, comparing and evaluating models is difficult because volatility cannot be directly observed. Thus, an additional factor is an approach to determining and calculating volatility. 

Stochastic volatility models are expressed as a stochastic process, which means that the volatility value at time t is latent and unobservable  \cite{ghysels19965}. On the other hand, some models try to estimate it as a non-parametric task, which implements more state freedom in parameter map conditions \cite{pagan1990alternative, west1995predictive}. However, non-parametric models did not show outperformance compared to more mainstream approaches.

%In \cite{pooter2008predicting, mcaleer2008realized} have been proposed evaluating stock market volatility by a smooth transition or threshold models. However, the log-likelihood of a sample may exhibit local maximums and require a significant amount of data to determine states and show low accuracy on an out-of-sample set \cite{pavlidis2012forecast}.

On the other hand, Neural Networks (NN) has been widely used for forecasting tasks, such as stock prices \cite{khan2011financial} or volatility predictions with additional input \cite{bucci2020realized}. Some other works focus on combination conditional volatility models with NN \cite {arneric2014garch}. This spectrum of work shows mixed performance results referring to volatility prediction tasks. In some cases were shown that non-parametric models, including NN, present poor forecasting performance for  an out-of-sample test \cite{clements1998comparison, pavlidis2012new}. In \cite{vortelinos2017forecasting} were shown that the feed-forward NN approximation is not well enough. Furthermore, \cite{vortelinos2017forecasting, bucci2017forecasting, miura2019artificial} demonstrate mixed forecasts accuracy on out-of-sample realized volatility by NN.
However, in \cite{rosa2014evolving, miura2019artificial}  were provided promising results for realized volatility forecasts out-of-sample accuracy. Implied volatility and realized volatility forecasting task were investigated in \cite{hamid2004using} was shown that NN performance as well as realized volatility. 

Also, in recent decades, there has been an increase in activity in finance using machine learning to detect anomalies and volatility prediction tasks with NLP analysis. 
%The advantage of machine learning is that we do not need to know the data structure and choose model parameters compere to GARCH models. 
The advantage of machine learning is that we do not need to specify the functional form apriori, but rather learn it from data.
However, these approaches have not shown substantial benefits in predicting realized volatility w.r.t. standard econometric volatility models. One possibility is that the asymmetrical influence of news on the volatility process ~\cite{black1976pricing, nelson1990stationarity} makes it particularly challenging to estimate the structure of the volatility process.

%volatility exhibiting asymmetrical behavior to news \cite{black1976pricing, nelson1990stationarity}, making it challenging to estimate volatility and define its underlying process. 

The problem with recurrent neural networks (RNN) is that distant memory of the first entries gradually disappears. After some iterations, the state of the RNN contains practically no traces of the first entries. For example, each long short-term memory (LSTM) cell defines what to remember, forget, and update memory using gateways. Thus, the LSTM network solves the problem of exploding or disappearing gradients. However, LSTM uses a large number of training parameters and therefore uses more resources. One of the difficulties is incorrectly choosing the optimal number of parameters when dealing with large sequences and caring about accuracy.
%Our work contributes to this literature. This work studies how Neural Networks (LSTM) can capture the underlying market volatility in proxy of realized volatility in finance and cryptocurrencies markets.
In this work, we studied and analyzed how Neural Networks (LSTM) can learn to capture the temporal structure of realized volatility.  
We aim to demonstrate how NN could be used for accurate volatility forecasting. For this purpose, we implement modern Neural Networks, Long Short Term Memory (LSTM) \cite{LSTM_hochreiter1997}, and the successor (next generation) of LSTM, Gated Recurrent Unit (GRU) \cite{GRU_cho2014}. Machine learning can approximate any linear and non-linear behavior and could learn data structure. However, hyper-parameters of NN could lead to different results. However, preprocessing of data and the combining procedure could be essential elements to achieve good performance. We investigated the approach with LSTM and GRU types for realized volatility forecasting tasks and compared the predictive ability of NN with widely used EWMA, HAR, GARCH-family models.

The remaining of the paper is organized as follows. Section 2 provides some preliminaries, such as the definition of realized volatility, showing some factors, for instance, data frequencies of time series.  Section 3 introduces all benchmarks and models for comparisons. Section 4 describes the NN architecture, data, methodology, and choice and reveals critical factors for such performance. Section 5 lists the analysis of forecast performance and evaluation for our models. Finally, the technical specification of all models is listed in Appendix.

\setlength{\belowdisplayskip}{0pt} \setlength{\belowdisplayshortskip}{0pt}
\setlength{\abovedisplayskip}{0pt} \setlength{\abovedisplayshortskip}{0pt}
\section{Preliminaries}

Simulations of stocks are often modeled using stochastic differential equations (SDE). Because of the randomness associated with stock price movements, the models cannot be developed using ordinary differential equations (ODE). A typical model used for stock price dynamics is the following stochastic differential equation:

\vspace{0.2cm}
\begin{equation}\label{d S}
\mathrm{d S}=\mu S d t+\sigma S d W_{t},
\end{equation}
\vspace{0.2cm}

where $S$ is the stock price, $\mu$ is the drift coefficient or mean of returns over some time period, $\sigma$ is the diffusion coefficient or the standard deviation of the same returns, and $W_{t}$ is known as the Wiener process or Brownian Motion.

Each increment $W_{i}$ is computed by multiplying a standard random variable $z_{i}$ from a normal distribution $N(0,1)$ with mean 0 and standard deviation 1 by the square root of the time increment $\sqrt{\Delta t_{i}}$:

\vspace{0.2cm}
\begin{equation}\label{W_{i}}
\mathrm{W_{i}}=z_{i} \sqrt{\Delta t_{i}}.
\end{equation}
\vspace{0.2cm}

The cumulative sum of the $W_{i}$ increments is the discretized path:

\vspace{0.2cm}
\begin{equation}\label{W_n_t}
\mathrm{W_{n}(t)} =\sum_{i=1}^{n} W_{i}(t)
\end{equation}
\vspace{0.2cm}

For the SDE  \autoref{W_n_t} with an initial condition for the stock price of $S(0)=S_{0}$, the closed-form solution of Geometric Brownian Motion (GBM) is

\vspace{0.2cm}
\begin{equation}\label{S(t)}
\mathrm{S(t)} = S_{0} e^{\left(\mu-\frac{1}{2} \sigma^{2}\right) t+\sigma W_{t}}
\end{equation}
\vspace{0.2cm}

Usually, stock prices volatility changes stochastically over time, but in GBM, volatility is assumed constant. In an attempt to make it practical as a model for stock prices, assumption that the volatility $\sigma$ is constant. A local volatility model could be derived from volatility as a deterministic function of the stock price and time. However,  we assume that the volatilities randomness is own and often described by a different equation driven by a different Winner process; the model is called a stochastic volatility model.
The Black-Scholes-Merton model assumes that volatility is constant. In practice, volatility varies through time, and a lognormal distribution of price will be obtained in the future \cite{hull1987pricing}. We could consider processes, for instance, retaining the property of continuous change in the price of an underlying asset but assuming a process other than geometric Brownian motion. Also, another alternative is to superimpose continuous changes in asset prices in leaps and bounds. Another alternative is to assume a process in which all asset price changes are jumped. The types of processes are collectively known as Levy processes~\cite{mandelbrot1967variation}. In general, we could also consider the Levy process, as implemented as a continuous-time stochastic process but with stationary independent increments. But let’s suppose first that the volatility parameter in the Geometric Brownian motion is a known function of time, the process followed by the asset price then:

\vspace{0.2cm}
\begin{equation}\label{dS}
\mathrm{dS} = \mu Sdt + \sigma(t) S W_t
%\mathrm{dS_{t}}=\mu S_{t}dt + \sigma S_{t}\dW_{t}\
\end{equation}
\vspace{0.2cm}

It was shown that \cite{hull1987pricing}, when volatility is stochastic but uncorrelated with the asset price, the price could be described by Black-Scholes-Merton integrated over the probability distribution of the average variance rate.
The case where the asset price and volatility are correlated is more complicated, and in this perspective, we should consider the model with two stochastic variables, price, and volatility.

In this paper, we will be focusing on the realized volatility of a particular asset. 
Let us denote the log-price of asset at time $t$ as $p(t)$. Furthermore, intra-day return with for a given segment $j$ and frequency $\Delta$ is  $r_{t-j*\Delta}=p_{t-j*\Delta}-p_{t-(j+1)*\Delta}$. Then, the daily realized volatility is defined as the square root of the sum of intra-day squared returns

\vspace{0.2cm}
\begin{equation}\label{RV}
\mathrm{RV_t^d} = \sqrt{\sum_{j=0}^{M-1} r^2_{t-j*\Delta}}
\end{equation}
\vspace{0.2cm}

Next, the realized volatility (RV) is the consistent estimator~\cite{barndorff2002econometric} of the squared root of the integrated variance (IV). 
There is even a more robust result ~\cite{barndorff2002estimating} stating that RV is a consistent estimator of quadratic variation if the underlying process is a semimartingale. The integrated one-day variance (IV) at day $t$ is defined as

\vspace{0.2cm}
\begin{equation}\label{IV_t^d}
\mathrm{IV_t^d} = \int_{t}^{t-1d} \sigma^2(\omega) d\omega
\end{equation}
\vspace{0.2cm}

The measurement is set by the specification for each case separately. For instance, the calculation could be done for realized volatility of daily returns over one month, or the realized volatility could be calculated on returns of minutes level per one day.  However, taking into account a time frame of less than 1-5 minutes could be complicated by noise \cite{Bollerslev_1986}. 

Underlying processes and parameters are time-varying. There could be various shapes of returns distributions \cite{Engle_1982, Bollerslev_1986, black1976pricing, nelson1990stationarity}. It may hamper the estimation of volatility and assessment of the underlying process. Consider $\sigma$ as a risk measure, which has drawbacks if we do not have a distribution or pricing dynamic. On the other hand, if we consider $\sigma$ as an uncertainty measure, we should consider a normal distribution for the returns \cite{poon2003forecasting}.
 
More investigated stylized volatility facts are fat tails, volatility clustering, leverage effects, long memory \cite{mandelbrot1967variation, fama1965behavior, black1976pricing, perry1982parkinson, ding1993long}. Stylized volatility facts have important implications for predicting volatility~\cite{diebold1998elements}. Given these patterns, volatility models have been designed in different ways.

\section{Models and Benchmarks}

This section aims to show various widely used models in measuring and forecasting volatility. We are focusing on models which primarily developed based on historical behavior and did not take into account additional input as news, market microstructure data, or macroeconomic rates \cite{atkins2018financial}.

We do not consider any classical non-parametric model, and it was demonstrated~\cite{pagan1990alternative, west1995predictive, poon2003forecasting} that non-parametric methods' accuracy is not high enough.

\subsection{GARCH-family models}

GARCH family models estimate historical volatility or conditional variance \cite{Bollerslev_1986}; the autoregressive conditional heteroskedasticity (ARCH)~\cite{Engle_1982} model is defined by

\vspace{0.2cm}
\begin{equation} \label{GARCH}
    \begin{cases}
        r_{t}=\mu+\sigma_{t} \varepsilon_{t} \\
        \sigma_{t}^{2}= \alpha_{0}+\sum_{i=1}^{q}\alpha_{i}\left(r_{t-i}-\mu\right)^{2}
    \end{cases}
\end{equation}
\vspace{0.2cm}

Where  $ \varepsilon_{t} \sim \mathcal{N}(0,1) \text { and } \sigma_{t} \varepsilon_{t} \sim \mathcal{N}\left(0, \sigma_{t}^{2}\right)$. Conditional variance refers to a linear function of the past conditional variance, but volatility is a deterministic function of historical returns.
In \cite{Bollerslev_1986} generalized the ARCH model by modeling the variance as an AR(p), GARCH-model estimates future volatility from a long-run variance, and recent variance estimation. \cite{Bollerslev_1986, mastro2014forecasting}.  Thus, the clustering effect is a sharp increase of volatility observed for some time and not followed by a sharp drop.

\vspace{0.2cm}
\begin{equation} \label{GARCH1}
\sigma_{\mathrm{t}}^{2}=\omega+\alpha * \mathrm{r}_{\mathrm{t}-1}^{2}+\beta * \sigma_{\mathrm{t}-1}^{2}, \text{   }    \omega>0, \alpha>0, \beta>0 \text{ and }  \gamma+\alpha+\beta=1
\end{equation}
\vspace{0.2cm}

Various extensions have been intruded since Exponential GARCH \cite{nelson1991conditional}, GJR-GARCH \cite{glosten1993relation}, and Threshold GARCH \cite{zakoian1994threshold} in this model encapsulating some of the stylized facts about volatility.  Among the multiple models of the GARCH family, the GARCH(1,1) model is the most popular~\cite{jafari2007does}. However, the complication and increase in the order of the model do not lead to more significant results but complicate the calculations \cite{hansen2005test}.

Although the GARCH process is driven by a single noise sequence, the diffusion is driven by two independent Brownian motions 
\vspace{0.2cm}
\begin{equation} \label{GARCH2}
    \left(W_{t}^{(1)}\right)_{t \geq 0} \text { and }\left(W_{t}^{(2)}\right)_{t \geq 0} 
\end{equation}
\vspace{0.2cm}

For example, the GARCH(1, 1) diffusion satisfies \cite{brockwell2006continuous}:
\vspace{0.2cm}
\begin{equation} \label{GARCH3}
    \begin{cases}
        d S_{t}=\sigma_{t} d W_{t}^{(1)} \\
        d \sigma_{t}^{2}=\theta\left(\gamma-\sigma_{t}^{2}\right)+\rho \sigma_{t}^{2} d W_{t}^{(2)}, \quad t \geq 0
    \end{cases}
\end{equation}
\vspace{0.2cm}

The behavior of this diffusion limit is therefore rather different from that
of the GARCH process itself since the volatility process $\left(\sigma_{t}^{2}\right)_{t \geq 0}$ evolves independently of the driving process $\left(W_{t}^{(1)}\right)_{t \geq 0}$ in the first of the equations.

\subsection{Heterogeneous Autoregression Realized Volatility}
The heterogeneous Autoregression Realized Volatility (HAR-RV)  model introduced by \cite{corsi2009simple} is based on the assumption that agents' behavior in financial markets, according to which they differ in their perception of volatility depending on their investment horizons and are divided into short-term, medium-term and long-term. The hypothesis of the existence of such heterogeneous structures in financial markets is based on the heterogeneous market hypothesis presented by~\cite{muller1993fractals}.

Different agents in a heterogeneous market have different investment periods and participate in trading on the exchange with different frequencies. The fundamental idea is that participants' basic decisions on different time horizons perceive and respond to different types of volatility. Suppose we assume that the memory of each component (the rate of decrease of the values of the autocorrelation function with an increasing lag) is exponentially decreasing with a particular time constant. In that case, the memory of the entire market will consist of many such exponentially decreasing components with various values of time constants. Moreover, their superposition will behave like a process with hyperbolically decreasing autocorrelation.

So, a short-term agent may react differently to fluctuations in volatility compared to a medium- or long-term investor. The HAR-RV model is an additive cascade of partial volatilities generated at different time horizons that follows an autoregressive process. Log realized volatilities in using the HAR-RV \cite{corsi2009simple} given the lognormal distribution. Thus, the HAR-RV approach is one more stable and accurate estimate for Realized Volatility \cite{corsi2012har}.
We worked with logs to avoid negativity issues and get approximately Normal distributions, consider the log $RV_{t}$ aggregated, as follows:

\vspace{0.2cm}
\begin{equation} \label{HAR}
\log \mathrm{RV}_{t}^{(n)}=\frac{1}{n}\left(\log \mathrm{RV}_{t}+\ldots+\log \mathrm{RV}_{t-n+1}\right)
\end{equation}
\vspace{0.2cm}

at the 3 different horizons, where $R V_{t}^{(d)}, R V_{t}^{(w)}$, and $R V_{t}^{(m)}$ are respectively the daily, weekly, and monthly observed realized volatilities.

\vspace{0.2cm}
\begin{equation} \label{HAR}
    \begin{cases}
        \tilde{\sigma}_{t+1 m}^{(m)}=c^{(m)}+\phi^{(m)} R V_{t}^{(m)}+\tilde{\omega}_{t+1 m}^{(m)} \\
        \tilde{\sigma}_{t+1 w}^{(w)}=c^{(w)}+\phi^{(w)} R V_{t}^{(w)}+\gamma^{(w)} \mathbb{E}_{t}\left[\tilde{\sigma}_{t+1 m}^{(m)}\right]+\tilde{\omega}_{t+1 w^{\prime}}^{(w)} \\
        \tilde{\sigma}_{t+1 d}^{(d)}=c^{(d)}+\phi^{(d)} R V_{t}^{(d)}+\gamma^{(d)} \mathbb{E}_{t}\left[\tilde{\sigma}_{t+1 w}^{(w)}\right]+\tilde{\omega}_{t+1 d}^{(d)}
    \end{cases}
\end{equation}
\vspace{0.2cm}

where $c^{(m)}$ - the constant and $\tilde{\omega}_{t+1 m}^{(m)}$ is an innovation that is simultaneously and consistently independent with a mean zero, for monthly aggregation. 
By straightforward recursive substitution of three factors Stochastic Volatility model where the factors are directly the past RV:
\vspace{0.2cm}
\begin{equation} \label{HAR}
\log \sigma_{t+1 d}^{(d)}=c+\beta^{(d)} \log \mathrm{RV}_{t}^{(d)}+\beta^{(w)} \log \mathrm{RV}_{t}^{(w)}+\beta^{(m)} \log \mathrm{RV}_{t}^{(m)}+\epsilon_{t+1 d}^{(d)}.
\end{equation}
\vspace{0.2cm}

And we could add a measure of errors $\epsilon$ for log RV:
\vspace{0.2cm}
\begin{equation} \label{HAR}
\log \sigma_{t+1 d}^{(d)}=\log \mathrm{RV}_{t+1 d}^{(d)}+\tilde{\epsilon}_{t+1 d}
\end{equation}
\vspace{0.2cm}

Therefore autoregression model in the RV with the feature of considering volatilities realized over different interval sizes:

\vspace{0.2cm}
\begin{equation} \label{HAR}
\log \mathrm{RV}_{t+1 d}^{(d)}=c+\beta^{(d)} \log \mathrm{RV}_{t}^{(d)}+\beta^{(w)} \log \mathrm{RV}_{t}^{(w)}+\beta^{(m)} \log \mathrm{RV}_{t}^{(m)}+\epsilon_{t+1 d}
\end{equation}
\vspace{0.2cm}

\subsection{Autoregressive Integrated Moving Average}
A feature of the time series for the Autoregressive Integrated Moving Average (ARIMA) application lies in the relationship between past values associated with current and future ones. Therefore, ARIMA models allow modeling integrated or differential-stationary time series. ARIMA's approach to time series is that the stationarity of the series is assessed first. Next, the series is transformed by taking the difference of the corresponding order, and some ARMA model is built for the transformed model \cite{box2015time}.

Box-Jenkins ARIMA (an integrated autoregressive moving average model) models are widely used for time series modeling. The models  are autoregressive models (AR), and moving average (MA) models. Let $y_{t}$ be a stationary time series that is the realization of a stochastic process. The general ARMA(p,q) model:

\vspace{0.2cm}
\begin{equation} \label{ARIMA}
y_{t}=\sum_{i=1}^{p} \phi_{i} y_{t-i}+a_{t}-\sum_{j=1}^{q} \theta_{j} a_{t-j}
\end{equation}
\vspace{0.2cm}

where $a_{t-j}$ is errors or residuals that are assumed to follow a normal distribution. When the time series exhibits nonstationary behavior, then the ARMA model can be extended and written using the differences:

\vspace{0.2cm}
\begin{equation} \label{ARIMA2}
W_{t}-\sum_{k=1}^{d} W_{t-k}=(1-B)^{d} Y_{t}
\end{equation}
\vspace{0.2cm}

where d is the order of the difference. Replacing the ARMA model with the differences in eq. gives us:

\vspace{0.2cm}
\begin{equation} \label{ARIMA2}
\phi_{p}(B)(1-B)^{d} Y_{t}=\theta_{q}(B) a_{t}
\end{equation}
\vspace{0.2cm}

the ARIMA model (p, d, q), where p is autoregressive order of AR(p), d - order of integration (differences of the original time series) and q order of the moving average MA(q).
The weakness of ARIMA models is the inability to model volatile variance. Since this type of variance is very common in currency pairs, persistent volatility cannot capture some of the basic properties of heteroscedastic volatility present in financial time series, such as stochastic volatility, clustering volatility, mean reversion, and fat tails.

\subsection{Moving average and exponential moving average}
Exponential volatility also represents random returns like a normal distribution. The peculiarity of this method of calculating volatility is that when calculating the standard deviation, the historical sample data are included in the calculation with weights that increase the weight of recent price movements in the sample compared to long-term actions. Exponentially Weighted Moving Average (EWMA) corrects simple variance problems by weighting individual periodic returns and giving more weight to recent observations. This method is a pretty good benchmark, along with a naive approach.

\vspace{0.2cm}
\begin{equation}\label{ewma}
\mathrm{E W M A}(\sigma_{n}^{2})=\alpha \sigma_{n-1}^{2}+(1-\alpha) R_{n-1}^{2}
\end{equation}
\vspace{0.2cm}

\subsection{LSTM and GRU}
A neural network could be imaged as a layered structure of connected neurons. Recurrent neural networks (RNN) are a class of NN that use previous outputs as inputs. Due to this, Recurrent neural networks could work with sequential data.
RNN neurons have cell state/memory, and input is processed according to this internal state, achieved by this recurrent mechanism. In RNNs, there are repeating activation modules of layers that allow them to store information, but not enough. However, RNNs often suffer from a vanishing gradient, which causes model training to become too slow or stop altogether. 

LSTM is a specific cell type in a recurrent neural network capable of catching long-term dependencies in data and fixing this exploding or vanishing gradient issue. Achievement is possible due to cell state and a combination of four gates interacting. The ability to eliminate or add information to the cell state, carefully regulated by gates structures, is a crucial difference with RNN. LSTM cells contain an additional state, which helps to internally maintain input memory, making them especially suitable for solving problems associated with sequential. Cell state conveys relative information along the entire chain of the sequence. The state of the cell reflects the corresponding information throughout the processing of the series, so data from earlier time steps can participate in later time steps \cite{colah2015}.

\vspace{0.2cm}
\begin{equation} \label{LSTM1}
f_{t}=\sigma\left(W_{f} \cdot\left[h_{t-1}, x_{t}\right]+b_{f}\right)
\end{equation}
\begin{equation} \label{LSTM1}
i_{t} =\sigma\left(W_{i} \cdot\left[h_{t-1}, x_{t}\right]+b_{i}\right)
\end{equation}
\begin{equation} \label{LSTM1}
\tilde{C}_{t} =\tanh \left(W_{C}
 \cdot\left[h_{t-1},x_{t}\right]+b_{C}\right)
\end{equation}
\begin{equation} \label{LSTM1}
C_{t}=f_{t} * C_{t-1}+i_{t} * \tilde{C}_{t}
\end{equation}
\begin{equation} \label{LSTM1}
o_{t} =\sigma\left(W_{o}\left[h_{t-1}, x_{t}\right]+b_{o}\right)
\end{equation}
\begin{equation} \label{LSTM1}
h_{t} = o_{t} * \tanh \left(C_{t}\right) 
\end{equation}
\vspace{0.2cm}

GRU is a new generation of recurrent neural networks, very similar to LSTMs, a variation on the LSTM, introduced in \cite{cho2014learning}.  The update gate combines the forget and input gates, and the cell state merges with the hidden state called the reset gate. As a result, the model is more straightforward, and the training procedure takes less time than the net with standard LSTM instead of GRU. By this modification, LSTM and GRU fix the vanishing/exploding gradient problem encountered by traditional RNN

\vspace{0.2cm}
\begin{equation} \label{LSTM1}
z_{t} =\sigma\left(W_{z} \cdot\left[h_{t-1}, x_{t}\right]\right) 
\end{equation}
\begin{equation} \label{LSTM1}
r_{t} =\sigma\left(W_{r} \cdot\left[h_{t-1}, x_{t}\right]\right) 
\end{equation}
\begin{equation} \label{LSTM1}
\tilde{h}_{t} =\tanh \left(W \cdot\left[r_{t} * h_{t-1}, x_{t}\right]\right) 
\end{equation}
\begin{equation} \label{LSTM1}
h_{t} =\left(1-z_{t}\right) * h_{t-1}+z_{t} * \tilde{h}_{t}
\end{equation}
\vspace{0.2cm}

Constructing a neural network is carried out by enumeration and finding the architecture with a minor error. In \cite{panchal2010searching} provided definitions of information criteria that can assist in defining a neural network. Nevertheless, learning different architectures and choosing by mistake remains the most commonly used.

%This paper examines the special cells for the RNN model, which is designed for sequential data. Also, we are going to optimize statistical time series model specifications for each experiment.

\section{Experiments}

\subsection{Accuracy measures}
The Mean Squared Error method minimizes the sum of the squared deviations of the actual values from the calculated sum of squared errors. If this sum is divided by the number of observations, we get MSE.

If the MSE did not have a power of 2, then the positive and negative deviations would cancel out, which would minimize the distance between the actual and calculated values. Thus, the presence of squares allows getting some estimate of the distance from the actual values to the line.

\vspace{0.2cm}
\begin{equation}\label{mse}
\mathrm{MSE}=\frac{1}{n} \sum_{i=1}^{n}\left(Y_{i}-\hat{Y}_{i}\right)^{2}
\end{equation}
\vspace{0.2cm}

For Mean Absolute Error, allow to get rid of signs and estimate the distance from the actual to the calculated values, which will need to be minimized. The undoubted advantage of MAE is that the modules do not multiply the deviations that are considered outliers. Therefore, this estimate is more robust than MSE.

\vspace{0.2cm}
\begin{equation}\label{mae}
\mathrm{MAE}=\frac{\sum_{i=1}^{n}\left|Y_{i}-\hat{Y}_{i}\right|}{n},
\end{equation}
\vspace{0.2cm}

where a test data sample has $n$ data points, $Y_i$ is the observed values of the predicted variable and  $\hat{Y}_i$ the predicted values. Also, we provide the result of the Root Mean Squared Error, which is just the square root of MSE; the Mean Absolute Percentage Error (MAPE) is measured percentage accuracy.

\subsection{Diebold-Mariano test}

The Diebold-Mariano test \cite{diebold2015comparing} compares the forecast accuracy of two forecast models.  In \cite{harvey1997testing} were addressed the finite sample properties of the Diebold-Mariano statistics. The additional assumption is that all autocovariance beyond some lag length is zero and modified test based on an approximately unbiased estimator.  The null hypothesis - the two models have equal forecast accuracy. However, the alternative hypothesis is that model 2 is less accurate than model 1. The alternative hypothesis is that model 2 is more accurate than model 1. Also, we could make two sides alternative hypotheses: method one and method 2 have different levels of accuracy.

\subsection{Value at Risk}
Value at Risk (VaR) \cite{jorion2000value} is a monetary estimate of the amount that the expected loss over a given period will not exceed with a given probability. VaR is the amount of losses on the investment portfolio for a fixed period if some adverse event occurs that may affect the market. As a time horizon, usually, one, five, or ten days is chosen since it is complicated to predict the behavior of the market for a more extended period. The level of acceptable risk is taken equal to 95 or 99 percent.

\subsection{Data}

This study analyzes how NN could estimate and predict realized volatility on different market structures, particularly indexes, stocks, and cryptocurrency. Also, we consider different time frames and analyze the effect and dependencies of data granularity \autoref{table:Data}.
% MAKE THE TABLE with all stat about data sets ?
%Citi
%The first time series is the dataset corresponding to a 1-minute time frame price of Citigroup Inc on NYSE, from 2 January 1998 till 28 April 2017 in total 2309304 price data points to calculate 4862 RV observation points. 

%Dell
%The second time series is the dataset corresponding to a 1-minute time frame price of Dell Inc from NYSE, from 2 January 1998 till 29 October 2013 in total 1730585 price data points to calculate 3982 days observation of RV.
%CBS
%The third time series is the dataset corresponding to a 1-minute time frame price of CBS Inc from NYSE, from 2 January 1998 till 25 October 2013 in total 1730411 price data points to calculate 3980 days observation of RV.
For \textbf{stock market data} for each day, we calculate RV based on the 1-minute time frame price observations. But returns are calculated on the last daily close price. As a result, our experimental data set is the table of intraday returns and corresponded Realized Volatility. The dataset was divided into three parts: training, validation, and test. The validation and the test sample of stocks are equivalent to 252 points that refer to the trading year.

%S&P500 
In the next part, we consider the\textbf{ S\&P~500 index}. We investigated the dataset corresponding to a daily time frame price of S\&P~500, in total 17923 price data points to calculate 815 RV observation points, calculation analogous to stocks. However, for this case, we calculated monthly Realized Volatility based on days log-returns instead of stocks data daily RV based on a 1-minute time frame. For the validation data set, and the test set we used 245 data points.

In the last part of our experiment, we investigated \textbf{cryptocurrency data}, particularly Bitcoin-USD and Ethereum-USD cryptocurrency pairs.
%Bitcoin
%We have prices of  Bitcoin corresponding to a 1-minute time frame, from 31 December 2011 till 22 April 2020, in total 4363457 price data points to calculate 3035 RV observation. 
%Ethereum
We had prices of  Bitcoin corresponding to a 1-minute time frame, however, Ethereum was taken from a high-frequency data frame corresponding to a seconds time frame. 
The cryptocurrency data was divided into three parts: training, validation, and test similar to stock and index data. The validation and the test sample are the exact sizes of the stock data set validation, and the test set 252 points.

%%%% Data table
%\vspace{1cm}
\begin{table}[ht]
\centering
\begin{tabular}{|l|l|l|l|l|l|l|l|}

\hline
Type of asset & Name &  Time frame & From & To &  Price points  & RV points & Aggregation \\
\hline

Stock & Citigroup Inc. & 1-minute & 02-Jan-98 & 28-Apr-17 & 2309304 & 4862 & day\\
Stock & Dell T. Inc.  & 1-minute & 02-Jan-98 &	29-Oct-13 & 1730585 & 3982 & day\\
Stock & CBS Inc. & 1-minute & 02-Jan-98 &	25-Oct-13 & 1730411 & 3980 & day\\
Index & S\&P500  & 1-day & 01-Feb-50 &	01-Dec-17 & 17923 & 815 & month\\
Cryptocurrency & Bitcoin-USD & 1-minute & 31-Dec-11 &	22-Apr-20 & 4363457 & 3035 & day\\
Cryptocurrency & Ethereum-USD & 1-second & 01-Feb-20 &	21-May-20 & 8237492 & 2650 & hour\\

\hline
\end{tabular}
\caption{\label{data}Description of the data for stock, index and cryptocurrency.}
\label{table:Data}
\end{table}
%%%%%

\subsection{Hyperparameter Optimization and Data Preprocessing}

%Neural networks could be complex, and there are no approaches to find the best-suited configuration for a particular task and saves computational resources.

We explored different configurations in terms of dynamic and objective outcomes, and it is a trial and test way to explore the best setups for a given task.
To find the best configuration of NN is necessary to conduct multiple experiments with different hyperparameters~\cite{panchal2010searching}. Therefore, we set up and interpreted the results of many training epochs and chose an appropriate one by the low \autoref{mse} and \autoref{mae} metrics for the validation data set.
In this paragraph, we list the hyperparameters of the experiments for LSTM. Layers and neurons: by implementing different settings we could archive specific types of estimation and deep structure.
Loss Function: MSE, MAE, and Huber - different loss functions could lift performance.
Activation function: determines how the weighted sum of the input is converted to output.
Batch size: the hyperparameter that refers to the number of training examples used per iteration.
Epochs: a number that determines how much interaction the learning algorithm runs throughout the learning process.
Optimizer: play a critical role in improving model accuracy. There were various options to be used, but the fastest optimizers do not usually achieve higher accuracy.
We performed  a standard LSTM model with hyperparameter optimization and introduced a window size optimization approach.

%%%% Hyperparameters  table 
\begin{table}[ht]
\centering
\begin{tabular}{|p{3cm}|p{6cm}|}
\hline
Hyperparameters  &  Values  \\  
\hline
Window size &  [1-50] \\
Layers  &  [1-5] \\
Units in layers &  [5,10,20,30,50,100,200] \\
Dropout &  [0.01, 0.05, 0.1, 0.2, 0.3] \\
Activation fn &  [linear, ReLU, SoftMAX, Than] \\
Loss fn &  [MAE, MSE, HUBER] \\
Epochs &  [1,2,3,4,5,10,20,30,50,100,1000] \\
Batch size &  [1,2,4,8,16,32,64,128] \\
Optimizer &  [RMSprop, SGD, ADAM] \\
\hline
\end{tabular}
\caption{\label{tab}The Table of LSTM Hyperparameters}
\label{table:hyper_tabel}
\end{table}
%%%%%

LSTM has a state layer $C_{t}$, practically $C_{t}$ helps LSTM capture long dependency besides length input. LSTM should be effective for any reasonable length. However, a correctly defined window length is a crucial parameter for this task to show high accuracy.
Worth noticing, if we look at HAR or ARIMA models, we have to define lag. How long the series contains the necessary data to forecast the next prediction horizon. However, it should be taken into account that when we differentiate price data, we vanish some of the data properties (since it is contained in the price) to obtain stationary series. Perhaps this is the critical feature why LSTM can achieve such results since a series with a length of n lags is  kept due to the cell memory mechanism \cite{laptev2017time}.
%approximated by several parameters~
We implemented LSTM with a high-level neural network API Keras, which is written in Python and running on top of TensorFlow~\cite{chollet2017deep}.

We used the following settings to train LSTM: we use as input $(RV_{t-n}, … , RV_{t-1})$ and $RV_{t}$ as target variable, we constructed our prediction task as an autoregression problem. In our experiment, autoregression showed more promising results than regression, which could be described as $RV_{t} \sim \alpha P_{t} + \epsilon_{t}$.

%%%, or other option to $(R_{t-n}, … , R_{t-1})$ and $RV_{t}$ – regression problem.
\begin{figure}[!tbp]
  \centering

  \begin{minipage}[b]{0.8\textwidth}
    \includegraphics[width = 1\linewidth]{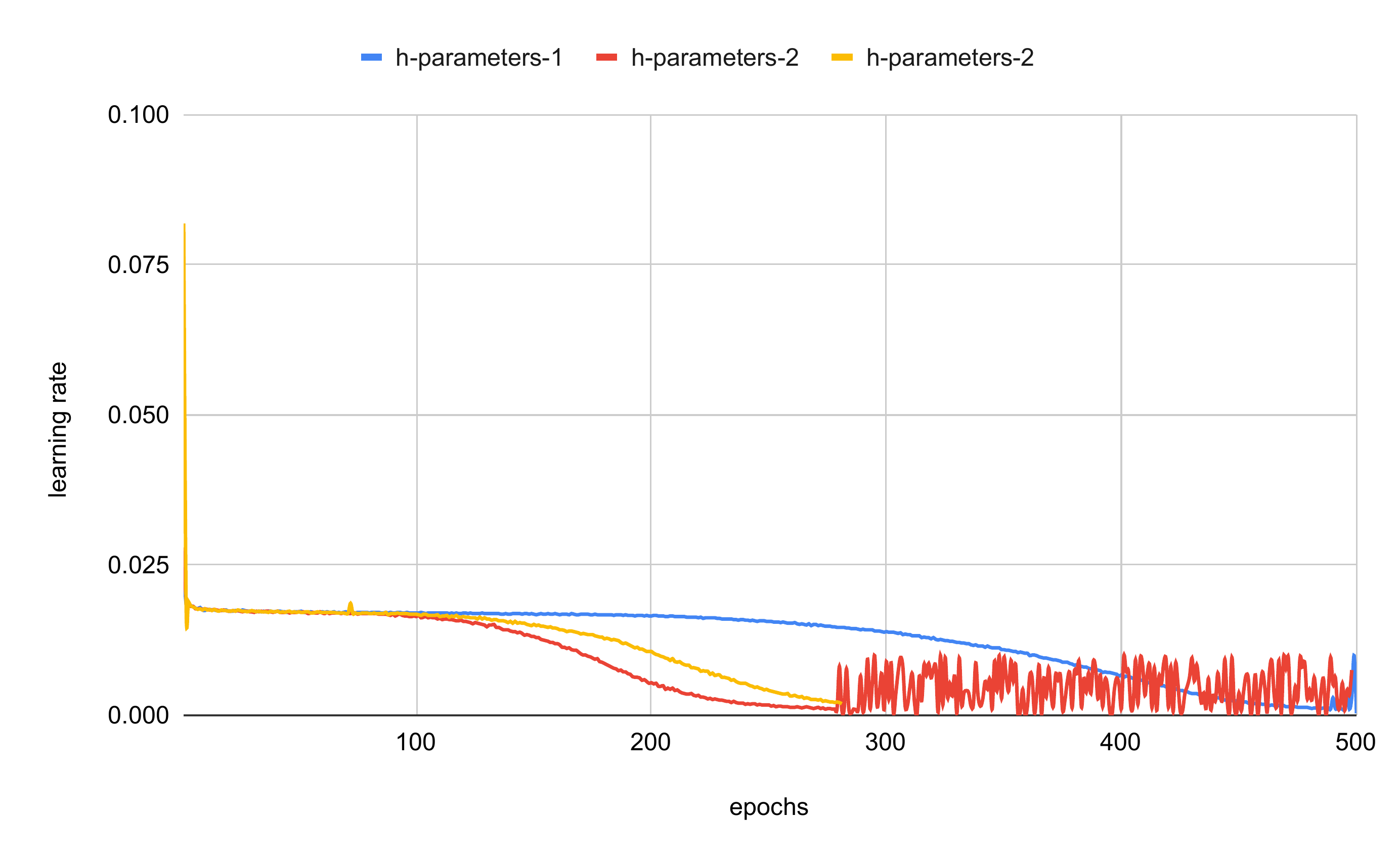}
    \caption{Learnign process of LSTM with different hyperptaratenes}
    \label{fig:learn_rate}
  \end{minipage}
\end{figure}

Another important hyperparameter is the activation function. There is a lack of experiments that consider different activation functions and performance for volatility prediction tasks. There are several possible variants we thought had more potential for our objective \autoref{table:hyper_tabel}. Moreover, another vital hyperparameter is an optimizer, commonly implemented ‘adam’ optimizer not always suited for all tasks. Nevertheless, as we have shown further, proper choosing functions and optimizers could make a significant difference.

However, the critical element in our experiment was dedicated to exploring the size of the sliding window. Therefore, we did not stop using a window size of n data points; we learned this hyperparameter by LSTM. We do not use augmented by external information approaches. 

Moreover, standardization is essential before training LSTM and can often improve the efficiency of training models. Therefore, we implemented an approach for normalizing input data from 0 to 1 by min-max scale.

We also optimized the parameter $\alpha$ of EWMA for MSE and MAE accuracy measures. As well we go through the optimization procedure for the HAR-RV model for MSE and MAE metrics. Nevertheless, we still have the standard model with parameters 1,5,22. In addition, ARIMA GARCH and GJR-GARCH were optimized based on information criteria, as were suggested in \cite{burnham2004multimodel}.

\section{Results}

%%%%%
In this empirical study, we sought to unravel the following questions. First, how do configurations of LSTM RNN (number of layers and neurons, activation functions, optimizer) network influence forecasting? 
Could hyperparameter optimized LSTM outperform the leading models in the field?
How much does the sliding window size affect LSTM efficiency and the model performance if this stage is excluded? 
%%%%%

In this section, we want to provide results for three different types of data sets we considered in our experiment: stocks, index, and cryptocurrency, and for three different aggregations of RV monthly, daily, and hourly.  We show the performance of the discussed models and detailed comparisons. We performed standard accuracy measures for the one-step-ahead prediction using RMSE, MAE metrics and tested the significance of improvement by the Diebold-Mariano test, \autoref{table:result_test} and the Appendix section.

The best results were demonstrated in the stock market by three types of models: LSTM window, HAR-RV, and EWMA. Interestingly, the LSTM window showed the best accuracy for MAE indicators at all validation samples of stocks. However, for RMSE, HAR hyperparameter optimization often became better, except for the Citi-stock data set, where EWMA was the best. Nevertheless, the out-of-sample RMSE and MAE accuracy-test LSTM with window optimized outperforms other models. DM-test also showed significant accuracy of the forecast quality compared to the closest competitor. According to the MAE out of sample test, the LSTM window optimized model was the best for the Dell stock data set, but HAR had the best RMSE without optimized parameters.

For performance comparison for different market conditions, we investigated the S\&P~500 index data set, where the data granularity does not have such extension as the data set of stocks or cryptocurrencies. Therefore, the results obtained should be inverse. On validation, the model could not show the best result, both for RMSE and MAE, although the gap was not significant at 0.1\text{\%} accuracy, and this was the second most accurate result. However, the LSTM with optimized window showed better accuracy for RMSE with a broad margin accuracy for the out-of-sample result. The DM test also showed the statistical significance of forecast accuracy. This is surprising enough, knowing about the intensity to train a network with a large number of parameters, and the need for large amounts of training data as opposed to econometric models.

LSTM also showed high accuracy performance in the cryptocurrency market, both on the validation and out of sample samples. However, it should be noted that the case of bitcoin out of sample HAR-RV was at the same level. We can trace the tendency that window optimization for LSTM is vital since with more granular data, the window increases. 

In the case of high-frequency data, the LSTM with the window-optimized approach shows more stable results. However, the LSTM with a fixed window experienced low-performance accuracy.

%%%% Dell table -test
\vspace{1cm}
\begin{table}[ht]
\centering

\begin{tabular}{|p{2.2cm}|p{1.25cm}|p{1.25cm}|p{1.25cm}|p{1.25cm}|p{1.25cm}|p{1.25cm}|p{1.25cm}|p{1.25cm}|p{1.25cm}|p{1.25cm}}
\hline
Model  & MSE \mbox{(e-05)}  &  RMSE \mbox{(e-03)}  &  MAE \mbox{(e-03)}  &  MAPE  &  DM test MSE  &  P-value MAE DM  &  DM test MAE  &  P-value MAE DM2  &  VaR \mbox{10 days}\\  
\hline
\end{tabular}

\begin{tabular}{p{17.25cm}}
\centering
\vspace{0.2cm}
Dell Technologies Inc stock - daily aggregation
\end{tabular}
%\noindent\rule{17.6cm}{0.4pt}
\begin{tabular}{|p{2.2cm}|p{1.25cm}|p{1.25cm}|p{1.25cm}|p{1.25cm}|p{1.25cm}|p{1.25cm}|p{1.25cm}|p{1.25cm}|p{1.25cm}|p{1.25cm}}
\hline
EWMA MSE        &  14,1142  &  11,8803  &  7,7360  &  43,3397  &   0,4295  &  0,6679    &  3,6767  &  0,0002    &  0,2544\\
EWMA MAE        &  13,8962  &  11,7882  &  7,5838  &  41,6260  &   0,1543  &  0,8774    &  3,2005  &  0,0015    &  0,2544\\
HAR MSE         &  13,7309  &  11,7178  &  8,2606  &  53,1560  &  -0,0507  &  0,9595    &  5,5384  &  $<$e-08   &  0,1533\\
HAR MAE         &  13,6397  &  11,6789  &  8,2126  &  52,9801  &  -0,1722  &  0,8633    &  5,5707  &  $<$e-08   &  0,1533\\
HAR             &  13,6397  &  11,6789  &  8,2126  &  52,9801  &  -0,1722  &  0,8633  &  5,5707  &  $<$e-08  &  0,1533\\
Last Values     &  17,9422  &  13,3948  &  7,6268  &  37,4586  &   2,3131  &  0,0215    &  1,7587  &  0,0798    &  0,1383\\
ARIMA           &  14,5067  &  12,0444  &  7,4324  &  40,7973  &   8,7439  &  $<$e-08   &  12,052  &  $<$e-08   &  0,1986\\
GARCH           &  79,9334  &  28,2725  &  13,720  &  82,4810  &   1,6996  &  0,0904    &  4,6902  &  $<$e-08   &  0,2040\\
GJR-GARCH       &  48,6597  &  22,0589  &  11,716  &  70,0866  &   1,4385  &  0,1515    &  4,1625  &  $<$e-08   &  0,1728\\
LSTM standart   &  19,9589  &  14,1276  &  9,0725  &  56,4351  &   4,7601  &  $<$e-08   &  6,7555  &  $<$e-08   &  0,1196\\
LSTM window     &  \textbf{13,0910}  &  11,4416  &  \textbf{6,8530}  &  35,8922  &  ***  &  ***  &  ***  &  ***  &  0,1280\\
\hline
\end{tabular}
\begin{tabular}{p{17.25cm}}
\centering
\vspace{0.2cm}
The Standard and Poor's 500 Index  (S\&P~500) - monthly aggregation
\end{tabular}
%\noindent\rule{17.6cm}{0.4pt}
\begin{tabular}{|p{2.2cm}|p{1.25cm}|p{1.25cm}|p{1.25cm}|p{1.25cm}|p{1.25cm}|p{1.25cm}|p{1.25cm}|p{1.25cm}|p{1.25cm}|p{1.25cm}}
\hline
EWMA MSE        &  51,7004 &  22,7377  &  14,8143  &  32,0655    &  3,4105 &  0,0007 &  4,3333     &  $<$e-08 &  0,2807\\
EWMA MAE        &  47,6968 &  21,8396  &  14,0323  &  30,1342     &  2,9263 &  0,0037 &  2,7277     &  0,0068 &  0,2805\\
HAR MSE         &  49,4088 &  22,2281  &  13,4078  &  28,1917    &  2,2882 &  0,0229 &  1,7231     &  0,0861 &  0,3185\\
HAR MAE         &  49,1897 &  22,1787  &  13,2569  &  27,3338    &  2,2819 &  0,0233 &  1,3088     &  0,1918  &  0,3177\\
HAR             &  45,0559 &  21,2263  &  \textbf{12,7762}          &  26,790 &  2,5911 &  0,0101     &  -0,3437 &  0,7313  &  0,3185\\
Last Values     &  43,1101 &  20,7629  &  14,2981   &  30,7739    &  0,2095 &  0,8341 &  2,2326     &  0,0264  &  0,3525\\
ARIMA           &  46,1339  &  21,4788  &  14,1858  &  30,8716    &  3,0891 &  0,0022  &  6,0096     &  $<$e-08        &  0,2823\\
GARCH           &  174,268  &  41,7454  &  34,1938  &  71,0463    &  9,5905 &  $<$e-08        &  18,972    &  $<$e-08        &  0,1040\\
GJR-GARCH       &  175,741  &  41,9215  &  34,5954  &  71,9429    &  9,6029 &  $<$e-08        &  19,613    &  $<$e-08        &  0,1046\\
LSTM standart    &  48,7022 &  23,5619  &  14,4471  &  32,9160    &  3,4729 &  $<$e-08        &  12,181     &  $<$e-08        &  0,1511\\
LSTM window     &  \textbf{41,6775}      &  20,4150  &  12,8474    &  28,505  &  ***      &  ***          &  ***      &  ***  &  0,3374\\
\hline
\end{tabular}
\begin{tabular}{p{17.25cm}}
\centering
\vspace{0.2cm}
Ethereum USD cryptocurrency - hourly aggregation 
\end{tabular}
%\noindent\rule{17.6cm}{0.4pt}
\begin{tabular}{|p{2.2cm}|p{1.25cm}|p{1.25cm}|p{1.25cm}|p{1.25cm}|p{1.25cm}|p{1.25cm}|p{1.25cm}|p{1.25cm}|p{1.25cm}|p{1.25cm}}
\hline
EWMA MSE        &  14,5095 &  12,0455    &  6,4798  &  26,0383    &  0,6170     &  0,5377 &  1,6013 &  0,1105     &  0,2472\\
EWMA MAE        &  15,1958 &  12,3271    &  6,6222  &  26,8891    &  0,5746     &  0,5660 &  1,8285 &  0,0686     &  0,2469\\
HAR MSE         &  13,1807 &  11,4807    &  6,6540  &  29,3306    &  -0,2493    &  0,8033 &  2,2370 &  0,0261     &  0,1787\\
HAR MAE         &  14,7789 &  12,1568    &  7,1052  &  31,6274    &  0,7861     &  0,4325 &  3,0047 &  0,0029     &  0,1813\\
HAR             &  \textbf{13,1785}      &  11,4797 &  6,6583    &  29,3342    &  -0,2573 &  0,7971 &  2,2608     &  0,0246  &  0,1787\\
Last Values     &  16,6921 &  12,9198    &  6,9548 &  28,7158    &  0,7582     &  0,4490 &  2,1630 &  0,0314      &  0,1683\\
ARIMA           &  15,1794 &  12,3204    &  7,3057 &  31,4821    &  2,7745     &  0,0059 &  7,1878 &  $<$e-08             &  0,1801\\
GARCH           &  584,465  &  76,4503    &  54,010 &  239,247    &  4,6473     &  $<$e-08  &  14,769  &  $<$e-08            &  0,5650\\
GJR-GARCH       &  513,573  &  71,6640    &  51,818 &  229,280   &  5,0685     &  $<$e-08  &  15,610 &  $<$e-08            &  0,5489\\
LSTM standart  &  27,5347 &  16,5935    &  10,357 &  37,0491    &  3,1206      &  0,0020 &  10,646  & $<$e-08            &  0,0946\\
LSTM window     &  13,5033 &  11,6203    &  \textbf{6,0079}  &  22,6919  &  ***  &  ***  &  ***  &  ***  &  0,1461\\
\hline
\end{tabular}
\caption{\label{tab}Dell-stock, S\&P~500 and ETH-USD out-of-sample, 1-step ahead accuracy test for 252 points.}
\label{table:result_test}
\end{table}
%%%%%

\section{Conclusion and discussion}
This work investigated whether neural networks can capture underlying processes for different markets such as stocks and cryptocurrency markets, gaining popularity in recent years. One of the peculiarities of RNN is that we do not need to know the parameters of the variable that we are trying to predict, which makes the processing data unified for the stock and cryptocurrency markets. This study used realized volatility as a target variable and formulated experiments as an autoregression problem.

We investigated the sets of hyperparameters for the model and their impact on performance. However, more crucial, we found out that the proper parameter of the rolling window in preprocessing procedure provides a better result in accuracy terms. Furthermore, we found that the more efficient size of the rolling window is from 5 to 12 periods. In this way, LSTM could outperform well-known models in this field, such as HAR-RV. Out-of-sample accuracy tests have shown that LSTM offers significant advantages in both types of markets. 

%The results also showed that the LSTM consistently outperformed the rival GRU, which is the successor to the LSTM model.

Despite the effectiveness of NN, there are still many difficulties, one of which, for example, is that they are black boxes and do not make it possible to analyze the relationships in the data. This work mainly examined what accuracy can be expected from LSTM in these tasks and what may influence the results. Another obvious problem is the number of parameters to be learned during the training process compared to HAR or EWMA models. On the other hand, modern computing power allows to compute it relatively fast.

\clearpage
\section{Appendix}

%%%% Citi table -valid
\vspace{1cm}
\begin{table}[ht]
\centering
\begin{tabular}{|l|l|l|l|l|l|l|}
%begin{tabular}{|p{2.6cm}|p{2cm}|p{2cm}|p{2cm}|p{2cm}|p{2cm}|p{2cm}|}
\hline
Model  &  Description & Parameters  & MSE \mbox{(e-05)}  &  RMSE \mbox{(e-03)}  &  MAE \mbox{(e-03)}  &  MAPE \\ 
\hline
EWMA MSE    &  alpha    &  0,55         & \textbf{5,3717} &  7,3292  &  4,8341  &  24,7917\\
EWMA MAE    &  alpha    &  0,39         &  5,4270  &  7,3668    &  4,8229   &  24,8157\\
HAR MSE     &  d,w,m:   &  2, 10, 110   &  5,4438  &  7,3782    &  4,9074   &  26,7506\\
HAR MAE     &  d,w,m:   &  2, 8, 110    &  5,5006  &  7,4166    &  4,8697   &  26,5904\\
HAR         &  d,w,m:   &  1, 5, 22     &  5,9750  &  7,7298    &  5,1681   &  28,5730\\
Last Values &  --       &  --           &  5,9519  &  7,7149    &  4,9908   &  25,9581\\
ARIMA       &  (p,d,q): &  (0, 1, 1)    &  5,5867  &  7,4744    &  4,8460   &  25,2255\\
GARCH       &  order    &  1,1          &  82,780  &  28,771    &  18,711   &  138,802\\
GJR-GARCH   &  order    &  1,1          &  148,29  &  38,509    &  19,945   &  138,565\\
LSTM standart   &  --       &  --           &  8,2386  &  9,0766    &  5,2918   &  22,7666\\
LSTM window &  window   &  7            &  5,3780  &  7,3334    &  \textbf{4,5678}  &  22,4754\\
\hline
\end{tabular}
\caption{\label{tab}Citi-stock, validation-dataset, 1-step ahead forecast, accuracy for 252 points.}
\label{table:Citi_valid}
\end{table}
%%%%%

%\begin{tabular}{|p{2.2cm}|p{1.25cm}|p{1.25cm}|p{1.25cm}|p{1.25cm}|p{1.25cm}|p{1.25cm}|p{1.25cm}|p{1.25cm}|p{1.25cm}|p{1.25cm}}

%%%% Citi table -test
\vspace{1cm}
\begin{table}[ht]
\centering
%\begin{tabular}{|l|l|l|l|l|l|l|l|l|l|l|}
\begin{tabular}{|p{2.2cm}|p{1.25cm}|p{1.25cm}|p{1.25cm}|p{1.25cm}|p{1.25cm}|p{1.25cm}|p{1.25cm}|p{1.25cm}|p{1.25cm}|p{1.25cm}}
\hline
%\thead{Model}   &  \thead{MSE \\ (e-05)}   &   \thead{ RMSE \\(e-03)}  &  \thead{ MAE \\(e-03)}  & \thead{ MAPE }  \\ \thead{DM test \\ mse } &  \thead{ P-value \\ mse DM } &\thead{  DM test \\ mae  } &  \thead{  P-value \\ mae DM}  & \thead{ VaR \\10 days} \\  

Model  & MSE \mbox{(e-05)}  &  RMSE \mbox{(e-03)}  &  MAE \mbox{(e-03)}  &  MAPE  &  DM test mse  &  P-value mae DM  &  DM test mae  &  P-value mae DM2  &  VaR 10 days\\ 
\hline
EWMA MSE        &  15,7334  &  12,5432  &  5,0453  &  26,0561  &  1,3554  &  0,1764  &  0,2452  &  0,8064  &  0,2210\\
EWMA MAE        &  14,4100  &  12,0041  &  5,0380  &  26,4374  &  1,6034  &  0,1100  &  0,3089  &  0,7576  &  0,2210\\
HAR MSE         &  12,8315  &  11,3276  &  5,1757  &  30,3151  &  0,8969  &  0,3706  &  1,6235  &  0,1057  &  0,1323\\
HAR MAE         &  12,8811  &  11,3495  &  5,2473  &  30,8161  &  1,0622  &  0,2891  &  2,2648  &  0,0243  &  0,1324\\
HAR             &  13,5329  &  11,6331  &  5,3237  &  30,5739  &  2,8817  &  0,0042  &  2,6992  &  0,0074  &  0,1313\\
Last Values     &  22,1741  &  14,8909  &  5,4230  &  27,4143  &  1,1727  &  0,2420  &  0,7796  &  0,4363  &  0,1237\\
ARIMA           &  16,5116  &  12,8497  &  5,2158  &  28,0231  &  4,7375  &  $<$e-08       &  7,7129  &  $<$e-08       &  0,1380\\
GARCH           &  111,245  &  33,3534  &  15,949  &  88,1419  &  3,1129  &  0,0020  &  6,3205  &  $<$e-08       &  0,2257\\
GJR-GARCH       &  177,384  &  42,1170  &  15,395  &  76,5359  &  2,5231  &  0,0122  &  4,3916  &  $<$e-08       &  0,2177\\
LSTM standart   &  16,2284  &  12,7390  &  5,9516  &  27,5748  &  3,3236  &  0,0010  &  3,2068  &  0,0015  &  0,0808\\
LSTM window     &  \textbf{12,4381}  &  11,1526  &  \textbf{4,9629}  &  28,6159  &  ***  &  ***  &  ***  &  ***  &  0,1291\\
\hline

\end{tabular}
\caption{\label{tab}Citi-stock, out-of-sample test, 1-step ahead forecast, accuracy for 252 points.}
\label{table:Citi_test}
\end{table}
%%%%%

\clearpage
%%%% Dell table -valid
\vspace{1cm}
\begin{table}[ht]
\centering
\begin{tabular}{|l|l|l|l|l|l|l|}
\hline
Model  &  Description & Parameters  & MSE \mbox{(e-05)}  &  RMSE \mbox{(e-03)}  &  MAE \mbox{(e-03)}  &  MAPE \\  
\hline
EWMA MSE        &  alpha    &  0,12         &  12,3047  &  11,0926  &  7,7043  &  30,4275\\
EWMA MAE        &  alpha    &  0,15         &  12,3367  &  11,1070  &  7,6868  &  30,1312\\
HAR MSE         &  d,w,m:   &  1, 12, 26    &  \textbf{12,0715}  &  10,9870  &  7,8279  &  31,8926\\
HAR MAE         &  d,w,m:   &  1, 11, 28    &  12,1427  &  11,0194  &  7,8179  &  31,8276\\
HAR             &  d,w,m:   &  1, 5, 22     &  12,3529  &  11,1143  &  7,9304  &  32,2523\\
Last Values     &  --       &  --           &  20,0029  &  14,1431  &  9,3039  &  34,3112\\
ARIMA           &  (p,d,q): &  (1, 1, 2)    &  13,2909  &  11,5286  &  8,1161  &  32,3217\\
GARCH           &  order    &  1,1          &  105,525  &  32,4846  &  23,036  &  89,3186\\
GJR-GARCH       &  order    &  1,1          &  105,525  &  32,4846  &  23,036  &  89,3186\\
LSTM standart      &  --       &  --           &  19,4981  &  13,9635  &  9,0714  &  27,9306\\
LSTM window     &  window   &  8            &  13,4469  &  11,5960  &  \textbf{6,9357}  &  23,2722\\
\hline
\end{tabular}
\caption{\label{tab}Dell-stock, validation-dataset, 1-step ahead forecast, accuracy for 252 points.}
\label{table:Dell_valid}
\end{table}
%%%%%

%%%% Dell table -test
\vspace{1cm}
\begin{table}[ht]
\centering
%\begin{tabular}{|l|l|l|l|l|l|l|l|l|l|l|}
\begin{tabular}{|p{2.2cm}|p{1.25cm}|p{1.25cm}|p{1.25cm}|p{1.25cm}|p{1.25cm}|p{1.25cm}|p{1.25cm}|p{1.25cm}|p{1.25cm}|p{1.25cm}}
\hline
Model  & MSE \mbox{(e-05)}  &  RMSE \mbox{(e-03)}  &  MAE \mbox{(e-03)}  &  MAPE  &  DM test mse  &  P-value mae DM  &  DM test mae  &  P-value mae DM2  &  VaR 10 days\\ 
\hline
EWMA MSE        &  14,1142  &  11,8803  &  7,7360  &  43,3397  &   0,4295  &  0,6679    &  3,6767  &  0,0002    &  0,2544\\
EWMA MAE        &  13,8962  &  11,7882  &  7,5838  &  41,6260  &   0,1543  &  0,8774    &  3,2005  &  0,0015    &  0,2544\\
HAR MSE         &  13,7309  &  11,7178  &  8,2606  &  53,1560  &  -0,0507  &  0,9595    &  5,5384  &  $<$e-08   &  0,1533\\
HAR MAE         &  13,6397  &  11,6789  &  8,2126  &  52,9801  &  -0,1722  &  0,8633    &  5,5707  &  $<$e-08   &  0,1533\\
HAR             &  13,6397  &  11,6789  &  8,2126  &  52,9801  &  -0,1722  &  0,8633  &  5,5707  &  $<$e-08  &  0,1533\\
Last Values     &  17,9422  &  13,3948  &  7,6268  &  37,4586  &   2,3131  &  0,0215    &  1,7587  &  0,0798    &  0,1383\\
ARIMA           &  14,5067  &  12,0444  &  7,4324  &  40,7973  &   8,7439  &  $<$e-08   &  12,052  &  $<$e-08   &  0,1986\\
GARCH           &  79,9334  &  28,2725  &  13,720  &  82,4810  &   1,6996  &  0,0904    &  4,6902  &  $<$e-08   &  0,2040\\
GJR-GARCH       &  48,6597  &  22,0589  &  11,716  &  70,0866  &   1,4385  &  0,1515    &  4,1625  &  $<$e-08   &  0,1728\\
LSTM standart   &  19,9589  &  14,1276  &  9,0725  &  56,4351  &   4,7601  &  $<$e-08   &  6,7555  &  $<$e-08   &  0,1196\\
LSTM window     &  \textbf{13,0910}  &  11,4416  &  \textbf{6,8530}  &  35,8922  &  ***  &  ***  &  ***  &  ***  &  0,1280\\
\hline
\end{tabular}
\caption{\label{tab}Dell-stock, out-of-sample test, 1-step ahead forecast, accuracy for 252 points.}
\label{table:Dell_test}
\end{table}
%%%%%

\clearpage
%%%% CBS table -valid
\vspace{1cm}
\begin{table}[ht]
\centering
\begin{tabular}{|l|l|l|l|l|l|l|}

\hline
Model  &  Description & Parameters  & MSE \mbox{(e-05)}  &  RMSE \mbox{(e-03)}  &  MAE \mbox{(e-03)}  &  MAPE \\  
\hline
EWMA MSE        &  alpha    &  0,47         &  5,2788           &  7,2655   &  4,3799         &  23,4852\\
EWMA MAE        &  alpha    &  0,68         &  5,3919           &  7,3429   &  4,3239         &  23,1228\\
HAR MSE         &  d,w,m:   &  1, 18, 110   &  \textbf{4,9344}  &  7,0245   &  4,5127         &  25,9045\\
HAR MAE         &  d,w,m:   &  1, 5, 90     &  4,9545           &  7,0388   &  4,4422         &  25,7425\\
HAR             &  d,w,m:   &  1, 5, 22     &  4,9786           &  7,0559   &  4,5514         &  26,3321\\
Last Values     &  --       &  --           &  5,9809           &  7,7336   &  4,4127         &  23,6452\\
ARIMA           &  (p,d,q): &  (2, 1, 1)    &  5,9117           &  7,6887   &  4,4555         &  24,1799\\
GARCH           &  order    &  1,1          &  39,787           &  19,946   &  14,889         &  117,575\\
GJR-GARCH       &  order    &  1,1          &  57,082           &  23,891   &  16,530         &  131,826\\
LSTM standart       &  --       &  --           &  5,5212           &  7,4304   &  4,9758         &  31,4461\\
LSTM window     &  window   &  5            &  5,1575           &  7,1816   & \textbf{4,1140} &  21,5335\\
\hline

\end{tabular}
\caption{\label{tab}CBS-stock, validation-dataset, 1-step ahead forecast, accuracy for 252 points.}
\label{table:CBS_valid}
\end{table}
%%%%%

%%%% CBS table -test
\vspace{1cm}
\begin{table}[ht]
\centering
\begin{tabular}{|p{2.2cm}|p{1.25cm}|p{1.25cm}|p{1.25cm}|p{1.25cm}|p{1.25cm}|p{1.25cm}|p{1.25cm}|p{1.25cm}|p{1.25cm}|p{1.25cm}}
\hline
Model  & MSE \mbox{(e-05)}  &  RMSE \mbox{(e-03)}  &  MAE \mbox{(e-03)}  &  MAPE  &  DM test mse  &  P-value mae DM  &  DM test mae  &  P-value mae DM2  &  VaR 10 days\\  
\hline
EWMA MSE        &  5,4257  &  7,3659     &  4,0731  &  28,2613  &  1,9592       &  0,0511     &  4,7255  &  $<$e-08    &  0,1823\\
EWMA MAE        &  6,0125   &  7,7540      &  4,2026  &  28,9532  &  2,0654      &  0,0399     &  4,2137  &  $<$e-08    &  0,1823\\
HAR MSE         &  4,9112  &  7,0080     &  4,2817  &  32,6479  &  0,4443       &  0,6571     &  6,9144  &  $<$e-08           &  0,1005\\
HAR MAE         &  4,9105  &  7,0075     &  4,2741  &  32,4164  &  0,4393      &  0,6607     &  6,3272  &  $<$e-08           &  0,1012\\
HAR             &  4,8683  &  6,9773     &  4,2497  &  32,2813  &  0,3093      &  0,7573     &  6,6897  &  $<$e-08           &  0,1006\\
Last Values     &  7,4220  &  8,6151     &  4,3838  &  30,3146  &  2,1572        &  0,0319      &  3,5024  &  0,0005    &  0,0935\\
ARIMA           &  6,1629   &  7,8504     &  4,4167  &  32,9735  &  3,9671      &  0,0000     &  8,6805  &  $<$e-08           &  0,1250\\
GARCH           &  14,151 &  11,895    &  10,280 &  92,2075  &  7,1467      &  $<$e-08            &  14,922 &  $<$e-08           &  0,1584\\
GJR-GARCH       &  15,325 &  12,379    &  10,436 &  92,8045  &  6,8824      &  $<$e-08            &  14,636  &  $<$e-08           &  0,1587\\
LSTM standart   &  5,0528  &  7,1083     &  4,6181  &  36,6197  &  0,7158      &  0,4747     &  7,8448  &  $<$e-08           &  0,1054\\
LSTM window     &  \textbf{4,7921}         &  6,9225  &  \textbf{3,2768} &  20,2920     &  ***  &  ***  &  ***   &  ***      &  0,0829\\
\hline
\end{tabular}
\caption{\label{tab}CBS-stock, out-of-sample test, 1-step ahead forecast, accuracy for 252 points.}
\label{table:CBS_test}
\end{table}
%%%%%

\clearpage
%%%% S&P table -valid
\vspace{1cm}
\begin{table}[ht]
\centering
\begin{tabular}{|l|l|l|l|l|l|l|}
\hline
Model  &  Description & Parameters  & MSE \mbox{(e-05)}  &  RMSE \mbox{(e-03)}  &  MAE \mbox{(e-03)}  &  MAPE \\  
\hline
EWMA MSE    &  alpha    &  0,2      &  41,8744             &  20,4632    &  9,3206     &  23,2058\\
EWMA MAE    &  alpha    &  0,28     &  42,2200             &  20,5475    &  9,2023     &  22,5636\\
HAR MSE     &  d,w,m:   &  3, 4, 55 &  \textbf{40,7741}    &  20,1926    &  8,5367     &  20,7522\\
HAR MAE     &  d,w,m:   &  3, 15, 30&  40,9492             &  20,2359    &  \textbf{8,4996}  &  20,4421\\
HAR         &  d,w,m:   &  1, 5, 22 &  45,3501             &  21,2955    &  8,9237     &  20,7829\\
Last Values &  --       &  --       &  64,2884             &  25,3551    &  10,872    &  25,7694\\
ARIMA       &  (p,d,q): &  (1, 1, 1)&  54,0468             &  23,2479    &  10,147     &  24,5666\\
GARCH       &  order    &  1,1      &  82,8823             &  28,7892    &  23,970    &  72,8169\\
GJR-GARCH   &  order    &  1,1      &  83,2874             &  28,8595    &  23,847    &  71,8954\\
LSTM standart   &  --       &  --       &  46,8581            &  22,2736    &  26,547     &  66,4353\\
LSTM window &  window   &  12       &  42,5604             &  20,6301    &  8,6655     &  20,6121\\
\hline

\end{tabular}
\caption{\label{tab}S\&P 500, validation-dataset, 1-step ahead forecast, accuracy for 245 points.}
\label{table:SandP_valid}
\end{table}
%%%%%

%%%% S%P table -test
\vspace{1cm}
\begin{table}[ht]
\centering
\begin{tabular}{|p{2.2cm}|p{1.25cm}|p{1.25cm}|p{1.25cm}|p{1.25cm}|p{1.25cm}|p{1.25cm}|p{1.25cm}|p{1.25cm}|p{1.25cm}|p{1.25cm}}

\hline

Model  & MSE \mbox{(e-05)}  &  RMSE \mbox{(e-03)}  &  MAE \mbox{(e-03)}  &  MAPE  &  DM test mse  &  P-value mae DM  &  DM test mae  &  P-value mae DM2  &  VaR 10 days\\  
\hline
EWMA MSE        &  51,7004 &  22,7377  &  14,8143  &  32,0655    &  3,4105 &  0,0007 &  4,3333     &  $<$e-08 &  0,2807\\
EWMA MAE        &  47,6968 &  21,8396  &  14,0323  &  30,1342     &  2,9263 &  0,0037 &  2,7277     &  0,0068 &  0,2805\\
HAR MSE         &  49,4088 &  22,2281  &  13,4078  &  28,1917    &  2,2882 &  0,0229 &  1,7231     &  0,0861 &  0,3185\\
HAR MAE         &  49,1897 &  22,1787  &  13,2569  &  27,3338    &  2,2819 &  0,0233 &  1,3088     &  0,1918  &  0,3177\\
HAR             &  45,0559 &  21,2263  &  \textbf{12,7762}          &  26,790 &  2,5911 &  0,0101     &  -0,3437 &  0,7313  &  0,3185\\
Last Values     &  43,1101 &  20,7629  &  14,2981   &  30,7739    &  0,2095 &  0,8341 &  2,2326     &  0,0264  &  0,3525\\
ARIMA           &  46,1339  &  21,4788  &  14,1858  &  30,8716    &  3,0891 &  0,0022  &  6,0096     &  $<$e-08        &  0,2823\\
GARCH           &  174,268  &  41,7454  &  34,1938  &  71,0463    &  9,5905 &  $<$e-08        &  18,972    &  $<$e-08        &  0,1040\\
GJR-GARCH       &  175,741  &  41,9215  &  34,5954  &  71,9429    &  9,6029 &  $<$e-08        &  19,613    &  $<$e-08        &  0,1046\\
LSTM standart    &  48,7022 &  23,5619  &  14,4471  &  32,9160    &  3,4729 &  $<$e-08        &  12,181     &  $<$e-08        &  0,1511\\
LSTM window     &  \textbf{41,6775}      &  20,4150  &  12,8474    &  28,505  &  ***      &  ***          &  ***      &  ***  &  0,3374\\
\hline
\end{tabular}
\caption{\label{tab}S\&P 500-index, out-of-sample test, 1-step ahead forecast, accuracy for 245 points.}
\label{table:SandP_test}
\end{table}
%%%%%

\clearpage
%%%% Bitcoin table -valid
\vspace{1cm}
\begin{table}[ht]
\centering
\begin{tabular}{|l|l|l|l|l|l|l|}

\hline
Model  &  Description & Parameters  & MSE \mbox{(e-05)}  &  RMSE \mbox{(e-03)}  &  MAE \mbox{(e-03)}  &  MAPE \\  
\hline
EWMA MSE    &  alpha    &  0,42     &  17,3251         &  13,1625    &  7,9259             &  21,6890\\
EWMA MAE    &  alpha    &  0,83     &  19,5003         &  13,9643    &  7,6130             &  21,1396\\
HAR MSE     &  d,w,m:   &  1, 5, 45 &  17,2831          &  13,1465    &  8,9144             &  29,9428\\
HAR MAE     &  d,w,m:   &  1, 4, 18 &  17,4985          &  13,2281    &  8,7833             &  29,1916\\
HAR         &  d,w,m:   &  1, 5, 22 &  17,4095         &  13,1945    &  8,8089             &  29,3820\\
Last Values &  --       &  --       &  21,3450         &  14,6099    &  7,8045             &  21,8513\\
ARIMA       &  (p,d,q): &  (3, 1, 3)&  25,0838         &  15,8378    &  8,5929             &  23,2321\\
GARCH       &  order    &  1,1      &  3348,44        &  182,987    &  145,84           &  718,887\\
GJR-GARCH   &  order    &  1,1      &  3115,92       &  176,519   &  137,70           &  687,079\\
LSTM standart  &  --       &  --       &  24,8242         &  15,7557    &  9,4235             &  23,9942\\
LSTM window &  window   &  8        & \textbf{16,3256} &  12,7771     &  \textbf{7,2293}    &  21,0681\\
\hline

\end{tabular}
\caption{\label{tab}Bitcoin-USD, validation-dataset, 1-step ahead forecast, accuracy for 252 points.}
\label{table:Bitcoin_valid}
\end{table}
%%%%%

%%%% BTCtable -test
\vspace{1cm}
\begin{table}[ht]
\centering
\begin{tabular}{|p{2.2cm}|p{1.25cm}|p{1.25cm}|p{1.25cm}|p{1.25cm}|p{1.25cm}|p{1.25cm}|p{1.25cm}|p{1.25cm}|p{1.25cm}|p{1.25cm}}
\hline

Model  & MSE \mbox{(e-05)}  &  RMSE \mbox{(e-03)}  &  MAE \mbox{(e-03)}  &  MAPE  &  DM test mse  &  P-value mae DM  &  DM test mae  &  P-value mae DM2  &  VaR 10 days\\  
\hline
EWMA MSE        &  27,7132 &  16,6472    &  8,3999     &  20,6981    &  1,4792     &  0,1403     &  1,8250     &  0,0691     &  0,3956\\
EWMA MAE        &  28,5206 &  16,8880    &  8,7378     &  21,7385    &  0,6404     &  0,5224     &  2,0189     &  0,0445     &  0,3956\\
HAR MSE         &  25,8854 &  16,0889    &  9,0112     &  25,4136    &  1,3917     &  0,1652     &  5,5401     &  $<$e-08            &  0,2805\\
HAR MAE         &  25,8716 &  16,0846    &  8,9371     &  25,1927    &  1,2148      &  0,2255     &  5,2337     &  $<$e-08            &  0,2793\\
HAR             &  26,1071 &  16,1576    &  8,9817     &  25,2597    &  1,6145     &  0,1076     &  5,6381     &  $<$e-08            &  0,2795\\
Last Values     &  29,8399 &  17,2742    &  8,9527     &  22,7036    &  0,6609     &  0,5092     &  2,0390     &  0,0424     &  0,2604\\
ARIMA           &  38,7652 &  19,6889    &  9,6716     &  23,0919    &  3,7090      &  0,0002     &  7,6181     &  $<$e-08            &  0,2598\\
GARCH           &  11810,2 &  343,660     &  176,07     &  403,906    &  3,8704     &  0,0001     &  9,2925     &  $<$e-08            &  1,5557\\
GJR-GARCH       &  13071,7 &  361,548     &  180,61     &  409,336    &  3,7308      &  0,0002     &  8,9830      &  $<$e-08            &  1,5891\\
LSTM standart   &  50,4766 &  22,4670    &  15,245    &  40,6017    &  4,0497     &  0,0001     &  13,873     &  $<$e-08            &  0,1493\\
LSTM window     &  \textbf{24,4282}        &  15,6290    &  \textbf{7,8785}  &  20,3574  &  ***  &  ***  &  ***  &  ***  &  0,2583\\
\hline
\end{tabular}
\caption{\label{tab}Bitcoin-USD, out-of-sample test, 1-step ahead forecast, accuracy for 252 points.}
\label{table:BTC_lstm}
\end{table}
%%%%%

\clearpage
%%%% ETH table -valid
\vspace{1cm}
\begin{table}[ht]
\centering
\begin{tabular}{|l|l|l|l|l|l|l|}

\hline
Model  &  Description & Parameters  & MSE \mbox{(e-05)}  &  RMSE \mbox{(e-03)}  &  MAE \mbox{(e-03)}  &  MAPE \\  
\hline
EWMA MSE        &  alpha    &  0,5      &  7,2411  &  8,5095     &  5,7709     &  22,5619\\
EWMA MAE        &  alpha    &  0,7      &  7,4220  &  8,6151     &  5,6981     &  22,3058\\
HAR MSE         &  d,w,m:   &  1, 5, 30 &  6,9555  &  8,3400    &  6,1236     &  26,7363\\
HAR MAE         &  d,w,m:   &  1, 4, 24 &  8,1742  &  9,0411     &  6,0497     &  22,9636\\
HAR             & d,w,m:    &  1, 5, 22 &  6,9657  &  8,3460     &  6,1170    &  26,7087\\
Last Values     &  --       &  --       &  8,2490  &  9,0824     &  5,9349     &  23,2643\\
ARIMA           &  (p,d,q): &  (1, 1, 2)&  7,4762  &  8,6465     &  5,7329     &  22,8089\\
GARCH           &  order    &  1,1      &  1003,9 &  100,19    &  60,207   &  284,042\\
GJR-GARCH       &  order    &  1,1      &  921,15 &  95,977     &  59,829      &  287,266\\
LSTM standart       &  --       &  --       &  24,642 &  15,698    &  11,812    &  40,9822\\
LSTM window     &  window   &  7        &  \textbf{6,7176} &  8,1961  &  \textbf{5,5248}  &  22,8914\\
\hline

\end{tabular}
\caption{\label{tab}Ethereum-USD, validation-dataset, 1-step ahead forecast, accuracy for 252 points.}
\label{table:Eth_valid}
\end{table}
%%%%%

%%%% ETH table -test
\vspace{1cm}
\begin{table}[ht]
\centering
\begin{tabular}{|p{2.2cm}|p{1.25cm}|p{1.25cm}|p{1.25cm}|p{1.25cm}|p{1.25cm}|p{1.25cm}|p{1.25cm}|p{1.25cm}|p{1.25cm}|p{1.25cm}}

\hline

Model  & MSE \mbox{(e-05)}  &  RMSE \mbox{(e-03)}  &  MAE \mbox{(e-03)}  &  MAPE  &  DM test mse  &  P-value mae DM  &  DM test mae  &  P-value mae DM2  &  VaR 10 days\\  
\hline
EWMA MSE        &  14,5095 &  12,0455    &  6,4798  &  26,0383    &  0,6170     &  0,5377 &  1,6013 &  0,1105     &  0,2472\\
EWMA MAE        &  15,1958 &  12,3271    &  6,6222  &  26,8891    &  0,5746     &  0,5660 &  1,8285 &  0,0686     &  0,2469\\
HAR MSE         &  13,1807 &  11,4807    &  6,6540  &  29,3306    &  -0,2493    &  0,8033 &  2,2370 &  0,0261     &  0,1787\\
HAR MAE         &  14,7789 &  12,1568    &  7,1052  &  31,6274    &  0,7861     &  0,4325 &  3,0047 &  0,0029     &  0,1813\\
HAR             &  \textbf{13,1785}      &  11,4797 &  6,6583    &  29,3342    &  -0,2573 &  0,7971 &  2,2608     &  0,0246  &  0,1787\\
Last Values     &  16,6921 &  12,9198    &  6,9548 &  28,7158    &  0,7582     &  0,4490 &  2,1630 &  0,0314      &  0,1683\\
ARIMA           &  15,1794 &  12,3204    &  7,3057 &  31,4821    &  2,7745     &  0,0059 &  7,1878 &  $<$e-08             &  0,1801\\
GARCH           &  584,465  &  76,4503    &  54,010 &  239,247    &  4,6473     &  $<$e-08  &  14,769  &  $<$e-08            &  0,5650\\
GJR-GARCH       &  513,573  &  71,6640    &  51,818 &  229,280   &  5,0685     &  $<$e-08  &  15,610 &  $<$e-08            &  0,5489\\
LSTM standart  &  27,5347 &  16,5935    &  10,357 &  37,0491    &  3,1206      &  0,0020 &  10,646  & $<$e-08            &  0,0946\\
LSTM window     &  13,5033 &  11,6203    &  \textbf{6,0079}  &  22,6919  &  ***  &  ***  &  ***  &  ***  &  0,1461\\
\hline

\end{tabular}
\caption{\label{tab}Ethereum-USD, out-of-sample test, 1-step ahead forecast, accuracy for 252 points.}
\label{table:Eth_lstm}
\end{table}
%%%%%

%%%%
\clearpage
\newpage
\begin{figure}[!tbp]
  \centering
  \begin{minipage}[b]{1\textwidth}
    \includegraphics[width=1\linewidth]{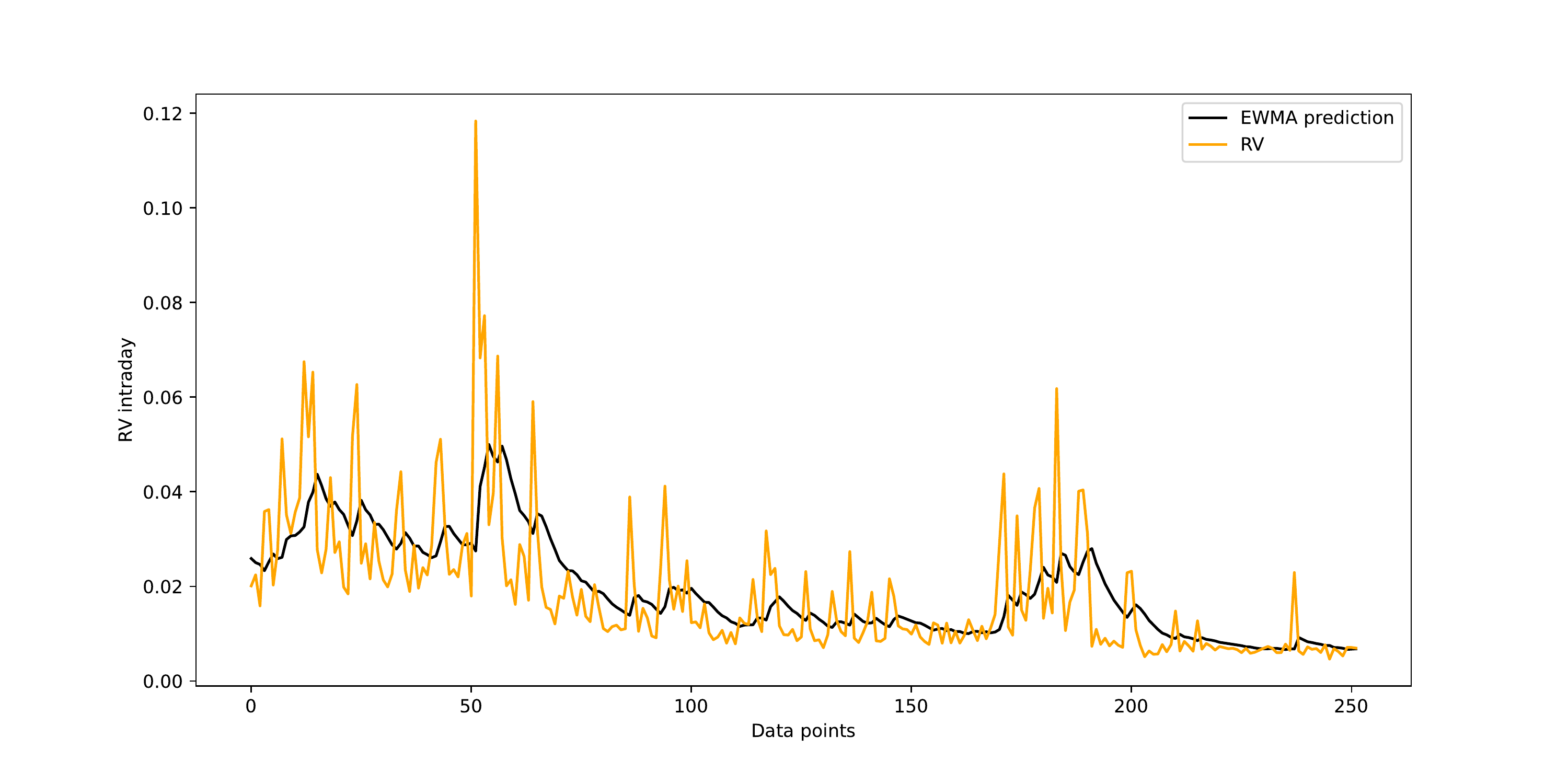}
    \caption{EWMA-model prediction for Dell stock}
    \label{fig:Dell_EWMA_252}
  \end{minipage}
  \hfill
  \begin{minipage}[b]{1\textwidth}
    \includegraphics[width = 1\linewidth]{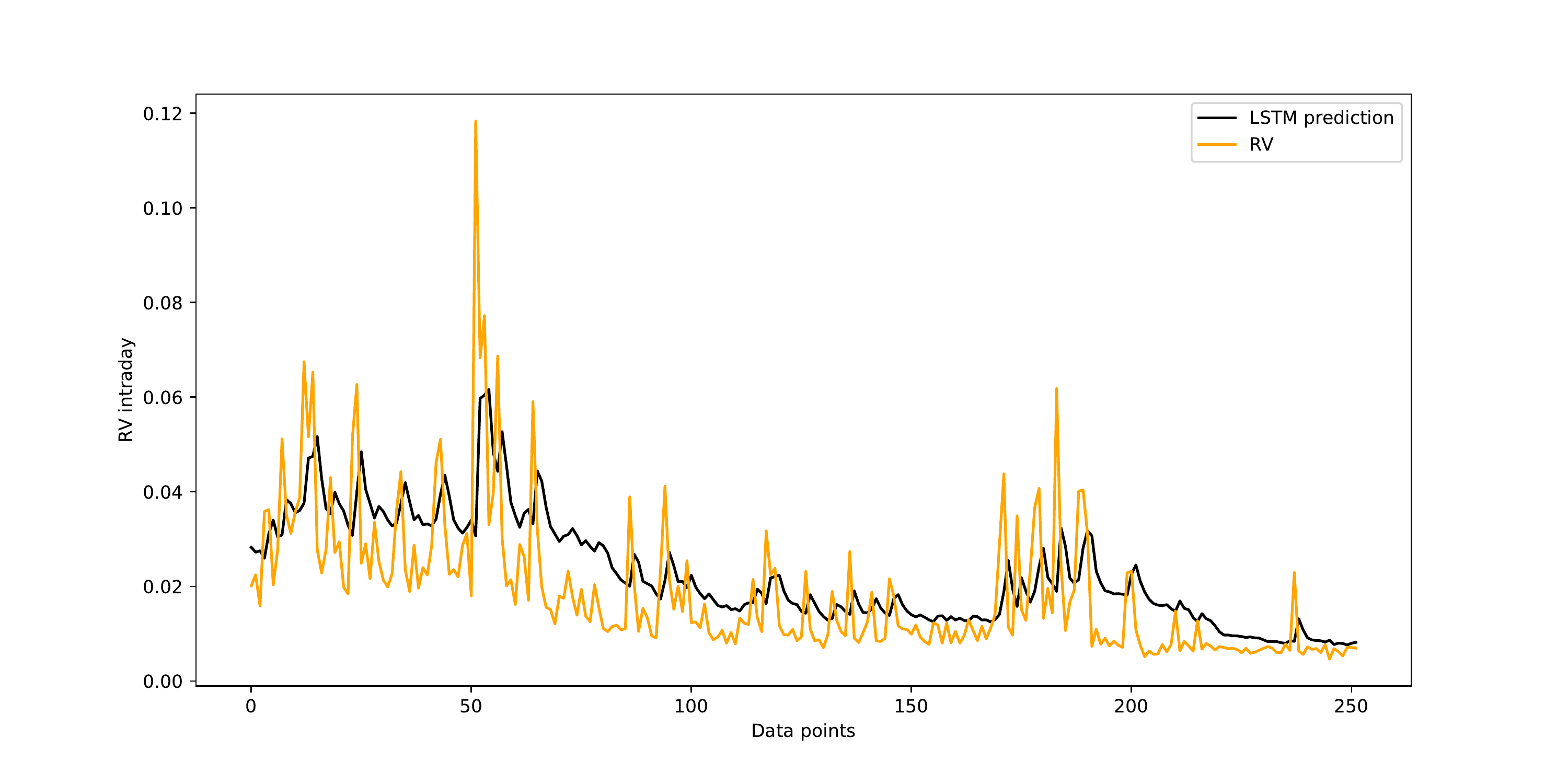}
    \caption{Long Short-Term Memory RNN prediction for Dell stock}
    \label{fig:Dell_LSTM_stand_252}
  \end{minipage}
\end{figure}

\begin{figure}[!tbp]
  \centering
  \begin{minipage}[b]{1\textwidth}
    \includegraphics[width=1\linewidth]{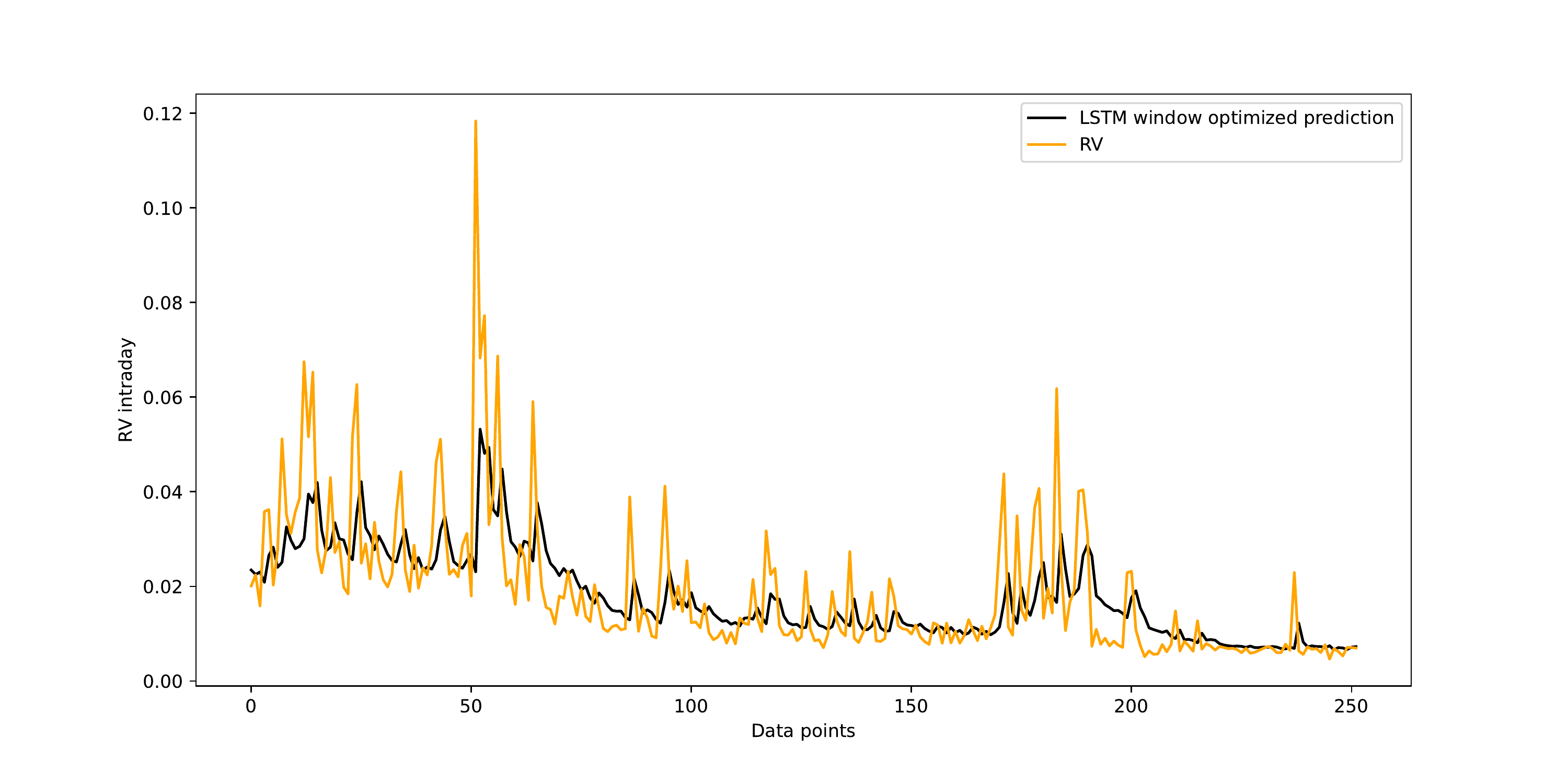}
    \caption{Long Short-Term Memory RNN with opt.window prediction for Dell stock}
    \label{fig:Dell_LSTM_window_252}
  \end{minipage}
  \hfill
  \begin{minipage}[b]{1\textwidth}
    \includegraphics[width = 1\linewidth]{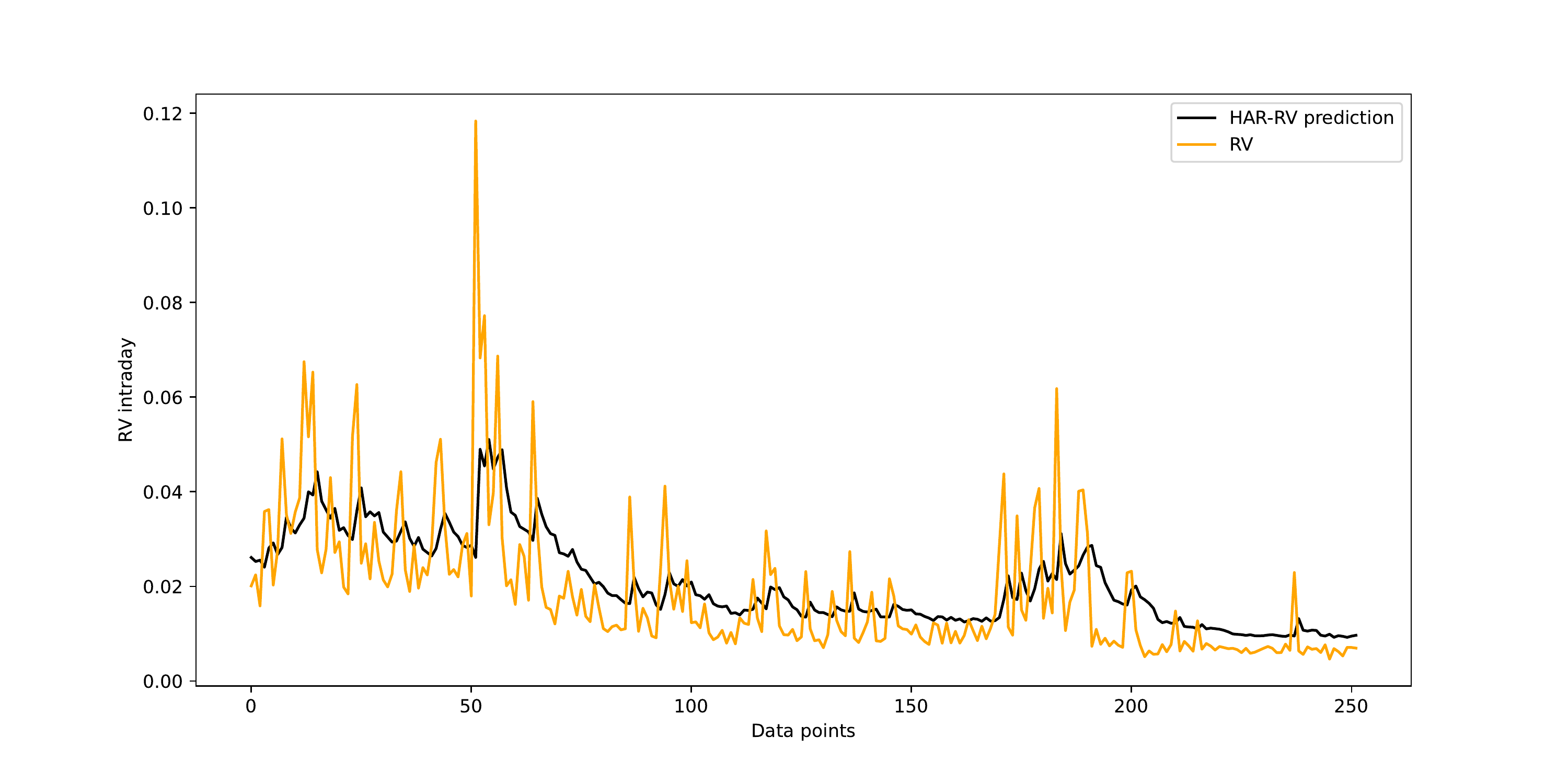}
    \caption{Heterogenous Autoregressive Realized Volatility Model prediction for Dell stock}
    \label{fig:Dell_HAR_252}
  \end{minipage}
\end{figure}
%%%%
%%%%
\clearpage
\newpage
\begin{figure}[!tbp]
  \centering
  \begin{minipage}[b]{1\textwidth}
    \includegraphics[width=1\linewidth]{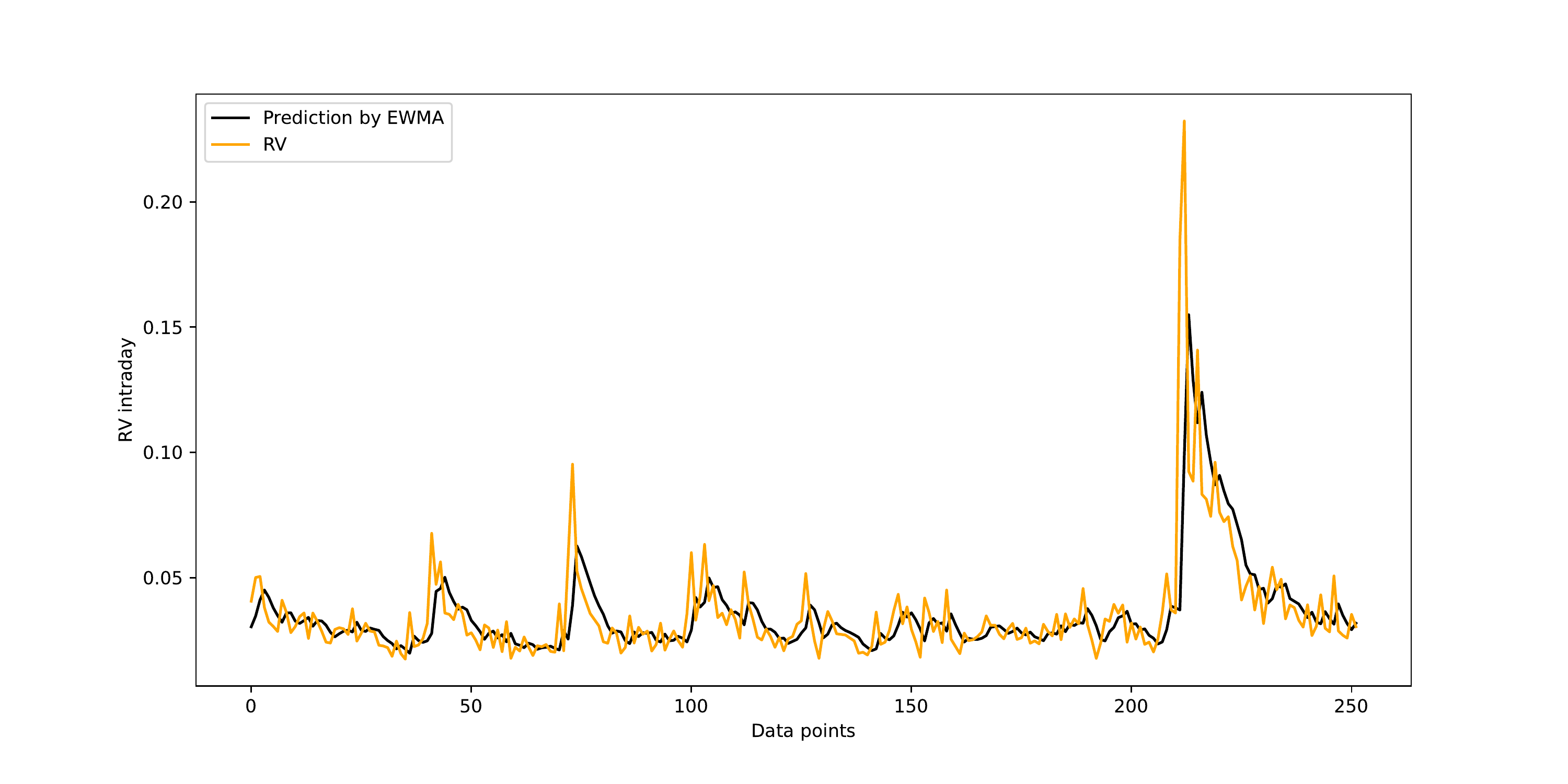}
    \caption{EWMA model prediction for Bitcoin cryptocurrency}
    \label{fig:Bitcoin_EWMA_252}
  \end{minipage}
  \hfill
  \begin{minipage}[b]{1\textwidth}
    \includegraphics[width = 1\linewidth]{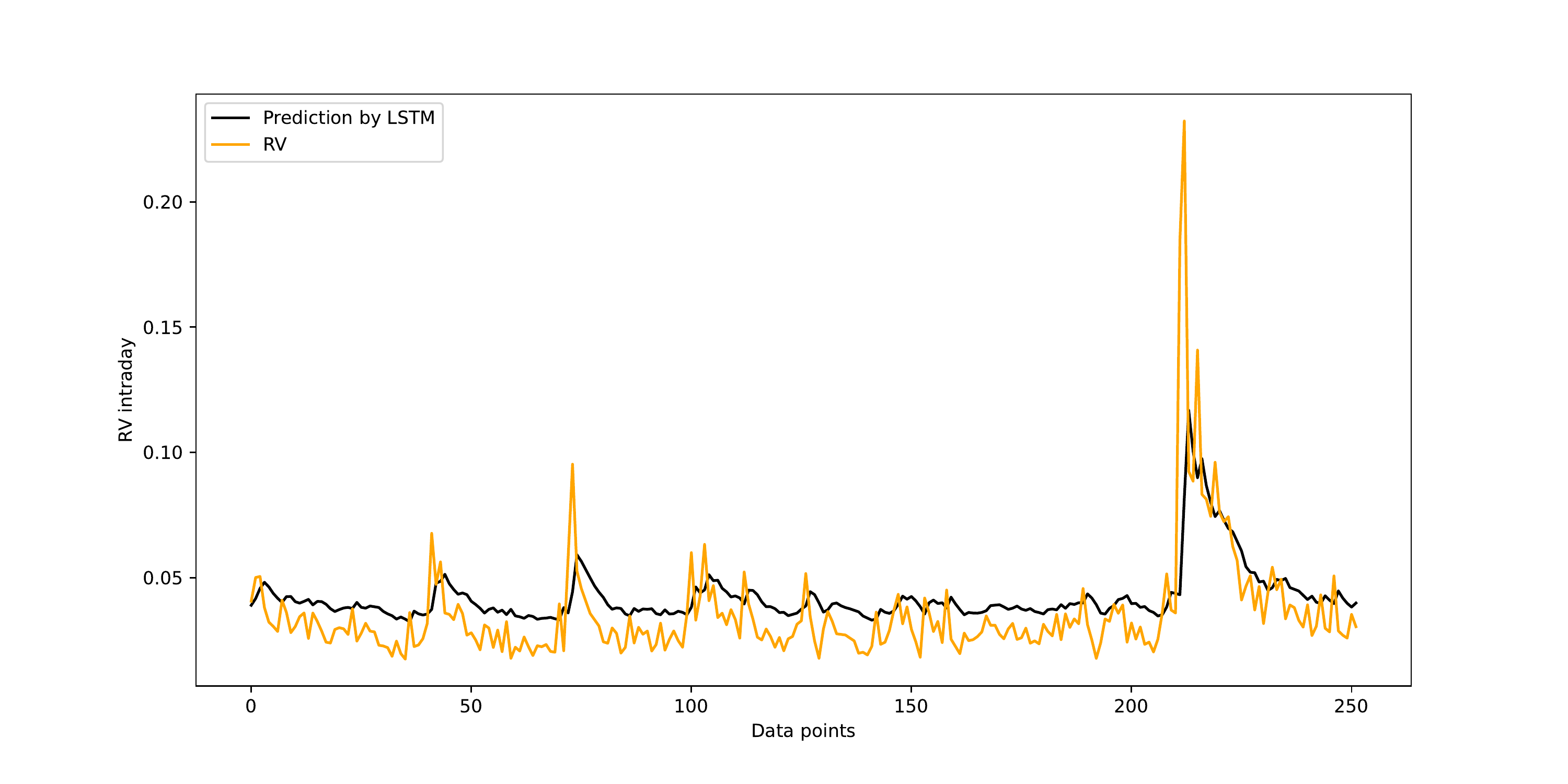}
    \caption{Long Short-Term Memory RNN prediction for Bitcoin cryptocurrency}
    \label{fig:Bitcoin_LSTM_stand_252}
  \end{minipage}
\end{figure}

\begin{figure}[!tbp]
  \centering
  \begin{minipage}[b]{1\textwidth}
    \includegraphics[width=1\linewidth]{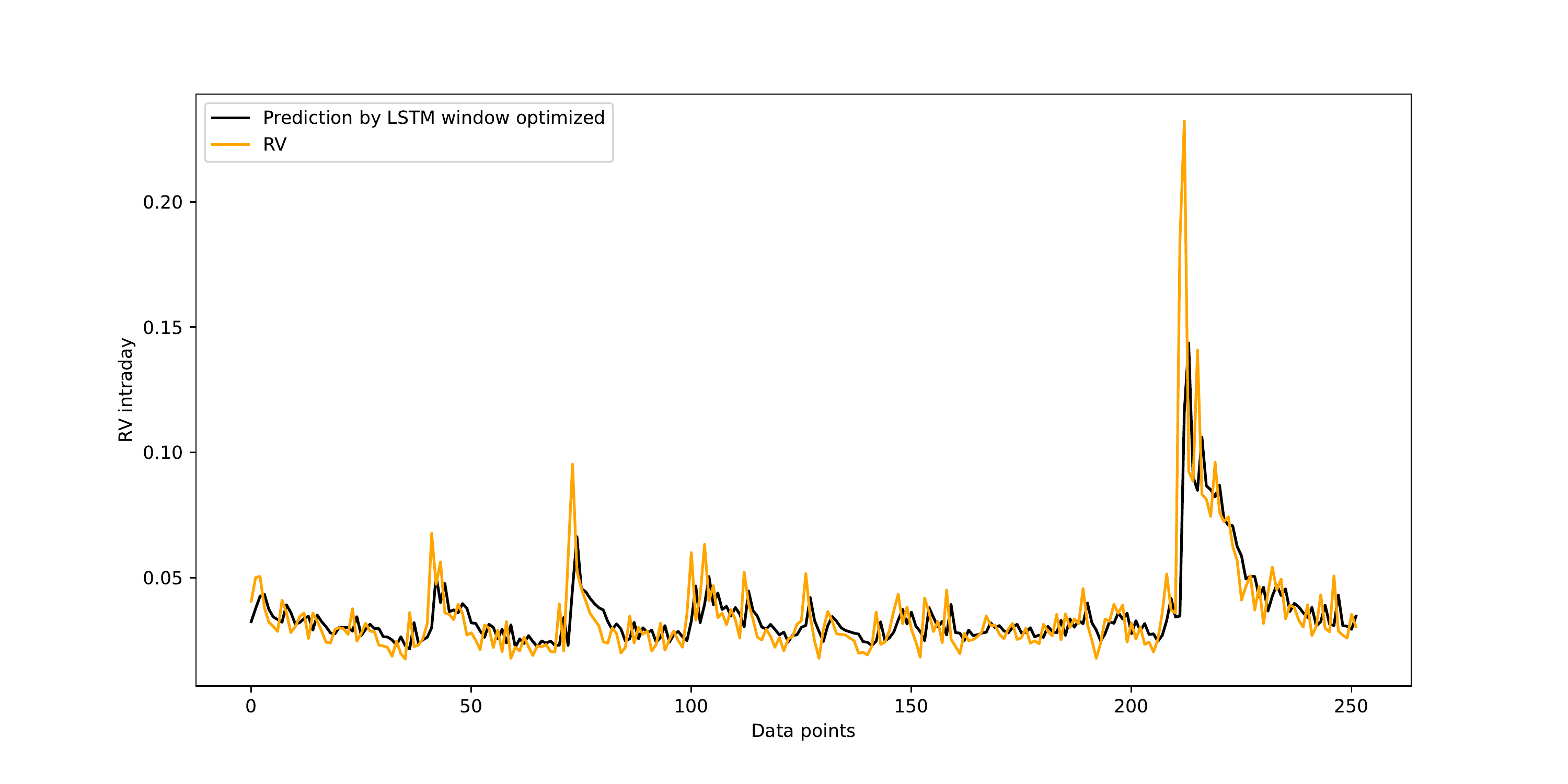}
    \caption{Long Short-Term Memory RNN with opt.window prediction for Bitcoin cryptocurrency}
    \label{fig:Bitcoin_LSTM_window_252}
  \end{minipage}
  \hfill
  \begin{minipage}[b]{1\textwidth}
    \includegraphics[width = 1\linewidth]{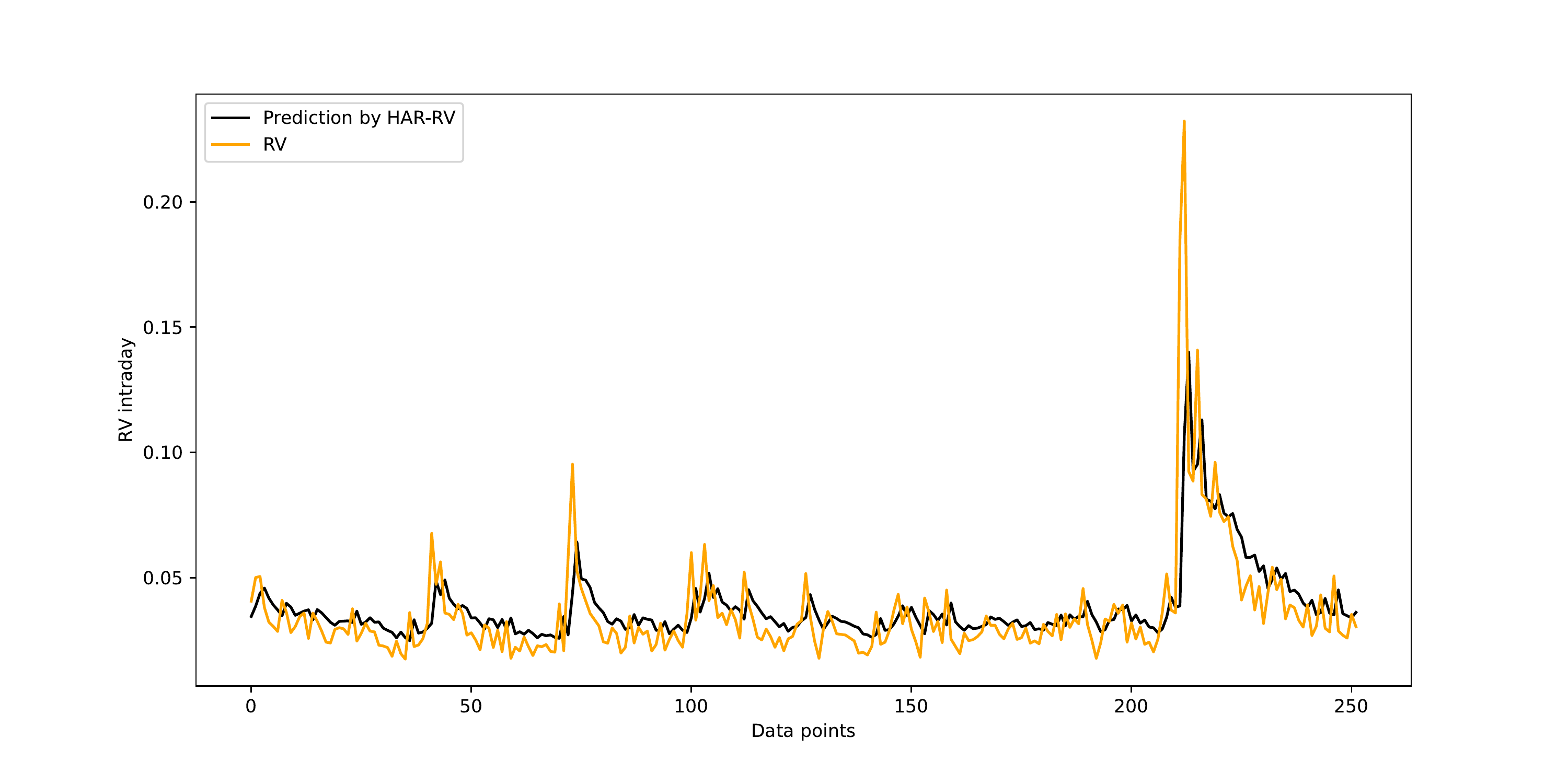}
    \caption{Heterogenous Autoregressive Realized Volatility Model prediction for Bitcoin cryptocurrency}
    \label{fig:Bitcoin_HAR_252}
  \end{minipage}
\end{figure}
%%%%%

\clearpage
\newpage
\bibliography{ref}

\end{document}